\documentclass[sigconf,nonacm]{acmart}

\setcopyright{none}
\acmDOI{}
\acmISBN{}
\renewcommand\footnotetextcopyrightpermission[1]{}

\usepackage[ruled,vlined]{algorithm2e}
\usepackage{graphicx}
\usepackage{textcomp}
\usepackage{xcolor}
\usepackage{listings}
\usepackage{array}
\usepackage{varwidth}
\usepackage{hyperref}
\usepackage{subcaption}
\usepackage{enumitem}

\usepackage[colorinlistoftodos]{todonotes} 
\usepackage{soul} % for the command \hl

\definecolor{mGreen}{rgb}{0,0.6,0}
\definecolor{mGray}{rgb}{0.5,0.5,0.5}
\definecolor{mPurple}{rgb}{0.58,0,0.82}
\definecolor{backgroundColour}{rgb}{0.95,0.95,0.92}
\definecolor{mygreen}{rgb}{0,0.6,0}
\definecolor{mygray}{rgb}{0.5,0.5,0.5}
\definecolor{mymauve}{rgb}{0.58,0,0.82}

\lstdefinestyle{CStyle}{
  backgroundcolor=\color{backgroundColour}, % choose the background color; you must add \usepackage{color} or \usepackage{xcolor}; should come as last argument
  breakatwhitespace=false,         % sets if automatic breaks should only happen at whitespace
  breaklines=true,                 % sets automatic line breaking
  captionpos=b,                    % sets the caption-position to bottom
  deletekeywords={...},            % if you want to delete keywords from the given language
  escapeinside={\%*}{*)},          % if you want to add LaTeX within your code
  extendedchars=true,              % lets you use non-ASCII characters; for 8-bits encodings only, does not work with UTF-8
  keepspaces=true,                 % keeps spaces in text, useful for keeping indentation of code (possibly needs columns=flexible)
  morekeywords={*,...},            % if you want to add more keywords to the set
  numbers=left,                    % where to put the line-numbers; possible values are (none, left, right)
  numbersep=5pt,                   % how far the line-numbers are from the code
  numberstyle=\tiny\color{mygray}, % the style that is used for the line-numbers
  rulecolor=\color{black},         % if not set, the frame-color may be changed on line-breaks within not-black text (e.g. comments (green here))
  showspaces=false,                % show spaces everywhere adding particular underscores; it overrides 'showstringspaces'
  showtabs=false,                  % show tabs within strings adding particular underscores
  tabsize=2,	                   % sets default tabsize to 2 spaces
  title=\lstname,                   % show the filename of files included with \lstinputlisting; also try caption instead of title
  belowcaptionskip=1\baselineskip,
  xleftmargin=\parindent,
  language=C,
  showstringspaces=false,
  basicstyle=\footnotesize\ttfamily,
  keywordstyle=\bfseries\color{green!40!black},
  commentstyle=\itshape\color{purple!40!black},
  identifierstyle=\color{blue},
  stringstyle=\color{orange}
}

\begin{document}

\title{An Irredundant and Compressed Data Layout to Optimize Bandwidth Utilization of FPGA Accelerators}

\author{Corentin Ferry}
\affiliation{
  \institution{Univ Rennes, CNRS, Inria, IRISA}
  \city{Rennes}
  \country{France}}
\email{cferry@mail.colostate.edu}

\author{Nicolas Derumigny}
\affiliation{
  \institution{Colorado State University, \\
  Univ. Grenoble Alpes, Inria, \\
  CNRS, Grenoble INP, LIG}
  \city{38000 Grenoble}
  \country{France}
}
\email{nicolas.derumigny@inria.fr}

\author{Steven Derrien}
\affiliation{
  \institution{Univ Rennes, CNRS, Inria, IRISA}
  \city{Rennes}
  \country{France}}
\email{Steven.Derrien@irisa.fr}

\author{Sanjay Rajopadhye}
\affiliation{
  \institution{Colorado State University}
  \city{Fort Collins, CO}
  \country{USA}}
\email{First.Last@colostate.edu}

\renewcommand{\shortauthors}{Ferry et al.}

\begin{abstract}

Memory bandwidth is known to be a performance bottleneck for FPGA accelerators, especially when they deal with large multi-dimensional data-sets.  A large body of work focuses on reducing of off-chip transfers, but few authors try to improve the efficiency of transfers. This paper addresses the later issue by proposing (i) a compiler-based approach to accelerator's data layout to maximize contiguous access to off-chip memory, and (ii) data packing and runtime compression techniques that take advantage of this layout to further improve memory performance.  We show that our approach can decrease the I/O cycles up
to $7\times$ compared to un-optimized memory accesses.
\end{abstract}


\maketitle

\section{Introduction}

FPGA accelerators have gained significant popularity in recent years, despite their inherent programming complexity.  High-Level Synthesis tools play a pivotal role in reducing the design challenges associated with FPGA acceleration, and facilitate their adoption in new application domains (e.g. machine learning).  % \todo{scientific computing is hardly a new domain}
% NICO: so let's go for ML >:-)

FPGA accelerator boards offer massive computational capabilities, but their performance is often hindered by an under-performing memory system, which becomes a performance bottleneck~\cite{lo2023hmlib}. This is especially true for accelerators that target compute intensive kernels operating on large data-sets. This issue is generally addressed through program transformations that increase temporal reuse, trading off-chip memory transfers for on-chip storage resource. 
% Although this approach helps reduce off-chip memory traffic, it does not necessarily improve off-chip bandwidth utilization. 
However, this approach does not primarily seek to optimize bandwidth usage (i.e. total amount of data transferred), leaving room for further improvement.
%contrarily to the technique presented in this paper.

Another approach consists in improving the effectiveness of the memory subsystem by reorganizing access patterns and data layout in order to exploit FPGA-specific constraints\cite{mayer2023employing}. 
%so as to better fit its characteristics. 
One classical way of doing so is by exploiting large burst-based transfers, which requires contiguous data in memory. This is however not easy when dealing with the multi-dimensional data-sets found in many applications, since the usual row-major and/or column major layouts only guarantee contiguity of data in only one dimension.

The problem becomes even more difficult when considering custom data formats (fixed/floating point) whose bitwidth do not correspond to the native memory bus interface. In such cases, the designer is left with two choices : padding the format to fit the bus width or deal with misaligned access, both choices incurring a loss of effective bandwidth.

Nevertheless, the ability to design application-specific hardware also brings opportunities to improve bandwidth efficiency. 
%For example, the multi-dimensional data sets accessed from external memory are often highly correlated, and could therefore be a good target for runtime-compression~\cite{}, further increasing the effectiveness of memory transfers.
For example, the fact that successive and/or ajacent values are numerically close (typical case in physical simulation) makes runtime compression a viable  strategy to increase the effectiveness of memory transfers.

In this paper, we present an automatic HLS optimisation flow that combines contiguity, data packing and
compression to maximize the utilization of bandwidth with custom data types. More precisely, our contributions are the following:
\begin{itemize}
    \item an algorithm to automatically derive (i) burst-friendly data layouts, and (ii) accelerator-specific access patterns that maximize contiguity while enabling data packing and compression,
    \item an automated code generation framework implementing the algorithm that generates synthesizable hardware,
    \item an evaluation of our approach on FPGA accelerators generated using the code generator that shows a up to $7\times$ decrease in I/O cycles.
\end{itemize}

This paper is organized as follows: Section~\ref{sec:background} presents the
concepts and core optimization techniques our work is relying on; 
Section~\ref{sec:layout} describes the memory layout transformation, and Section~\ref{sec:implementation} explains how we automatically apply it
to FPGA accelerators. Finally, Section \ref{sec:evaluation} validates our
approach and discusses it on a series of benchmarks.

\section{Background}
\label{sec:background}

% \subsection{Illustrative example: 1D Jacobi stencil}
% \label{sec:running-example}

% To illustrate the flow proposed in this paper, we use a a one-dimensional 
% Jacobi stencil as running example. This kernel updates a one-dimensional sequence
% of values, and computes each point as a weighted average of it and
% its neighbors:

% $$ c_{t+1, i} = \frac{1}{3}\left(c_{t, i-1} + c_{t, i} + c_{t, i+1}\right) $$

% A C implementation of this stencil is provided in the PolyBench/C suite
% as the following code:
% % Please do not cut code...

% \noindent\begin{minipage}{\linewidth}
% \begin{lstlisting}[style=CStyle, belowskip=-2 \baselineskip]
% for (t = 0; t < _PB_TSTEPS; t++) {
%     for (i = 1; i < _PB_N - 1; i++)
%         B[i] = 0.33 * (A[i-1] + A[i] + A[i+1]);
%     for (i = 1; i < _PB_N - 1; i++)
%         A[i] = 0.33 * (B[i-1] + B[i] + B[i+1]);
% }
% \end{lstlisting}
% \end{minipage}

% This stencil operates over a two-dimensional iteration domain $time \times space$ where each point 
% has a coordinate $(t, i)$. Because such a domain may be arbitrarily 
% large, the whole dataset may not fit into FPGA on-chip memory, and needs to 
% be optimized before is can be mapped to the FPGA. 

\subsection{Locality optimizations}
\label{sec:locality-opt}
As manually optimizing an HLS design at the source level is a tedious process, automated approaches are now routinely used for HLS/FPGA targets in order to exploit parallelism and locality at multiple levels, which are often implemented as source-to-source compilers~\cite{liu2017polyhedral, liu2015offline, cong2016source, ye2022scalehls}. Due to the inherent regularity of HLS-valid code, loop transformations engines such as PolyOpt/HLS~\cite{Pouchet_2013} and POLSCA~\cite{zhao2022polsca} excel at this task. However, such transformations are calibrated to improve computation time of benchmarks and do not seek to change the memory layout to enforce memory access contiguity.

\begin{figure}
    \centering
    \includegraphics[width=\columnwidth]{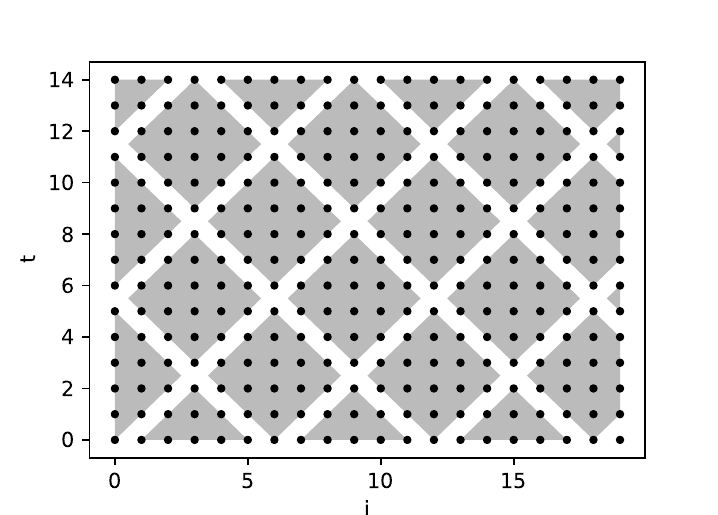}
    \caption{Domain of the Jacobi stencil divided into tiles of size $6 \times 6$.
    Each tile contains 18 $(t, i)$ points corresponding to 18 computations of $c_{t, i}$s.}
    \label{fig:jacobi1d-tiles}
\end{figure}

\begin{figure}
    \centering
    \includegraphics[width=\columnwidth]{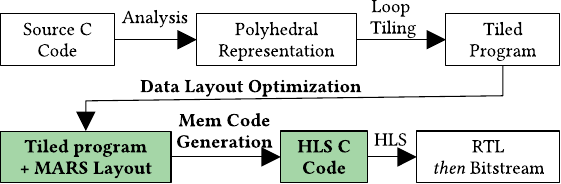}
    \caption{Compiler flow (our contributions in green)}
    \label{fig:polyhedral-flow}
\end{figure}

\label{sec:loop-tiling}

Indeed, to get the best runtime performance from an FPGA accelerator, it is necessary
to limit its off-chip memory accesses as much as possible. Only a fraction of 
large data sets can fit the limited size of on-chip memory; programs operating 
on large data sets must therefore be transformed to work on smaller workloads at a time.
Loop tiling does this: it breaks large spaces into smaller sub-problems called
\emph{tiles}, where the on-chip memory and parallelism requirements of each 
tile match those available on the chip.

An accelerator for a tiled program processes the domain tile by tile. 
To execute a tile, the accelerator needs to retrieve intermediate 
results from previously executed tiles. 
These intermediate results are located outside of the accelerator, 
in off-chip memory, and need to be copied into on-chip memory.

The amount of on-chip memory needed to run a tile is directly 
influenced by the tile's shape and size. In addition, when the tile 
size increases, the overall off-chip memory access is reduced, thus 
improving the overall arithmetic Intensity.
Selection of the best tile shape and size is outside the scope of this
work, and mainly depends on the performance / area trade-off desired by the designer.

\subsection{Illustrative example: 1D Jacobi stencil}
\label{sec:running-example}

To illustrate the flow proposed in this paper, we propose a
Jacobi-1D stencil as running example. This kernel updates a one-dimensional sequence
of values, and computes each point as a weighted average of it and
its neighbors:

$$ c_{t+1, i} = \frac{1}{3}\left(c_{t, i-1} + c_{t, i} + c_{t, i+1}\right) $$

A C implementation of this stencil is provided in the PolyBench/C suite
as the following code:
% Please do not cut code...

\noindent\begin{minipage}{\linewidth}
\begin{lstlisting}[style=CStyle, belowskip=-2 \baselineskip]
for (t = 0; t < _PB_TSTEPS; t++) {
    for (i = 1; i < _PB_N - 1; i++)
        B[i] = 0.33 * (A[i-1] + A[i] + A[i+1]);
    for (i = 1; i < _PB_N - 1; i++)
        A[i] = 0.33 * (B[i-1] + B[i] + B[i+1]);
}
\end{lstlisting}
\end{minipage}

This stencil operates over a two-dimensional iteration domain $time \times space$ where each point 
has a coordinate $(t, i)$. Because such a domain may be arbitrarily 
large, the whole dataset may not fit into FPGA on-chip memory, and needs to 
be optimized before it can be mapped to the FPGA. As stated in the previous subsection, 
this naive implementation of Jacobi-1D cannot fit on-chip for gigabyte-scale problem sizes, thus requiring tiling. For the sake of simplicity of the illustration, we have chosen small, diamond-shaped tiles, illustrated in Figure~\ref{fig:jacobi1d-tiles}.

\begin{figure}
    \centering
    \includegraphics[width=\columnwidth]{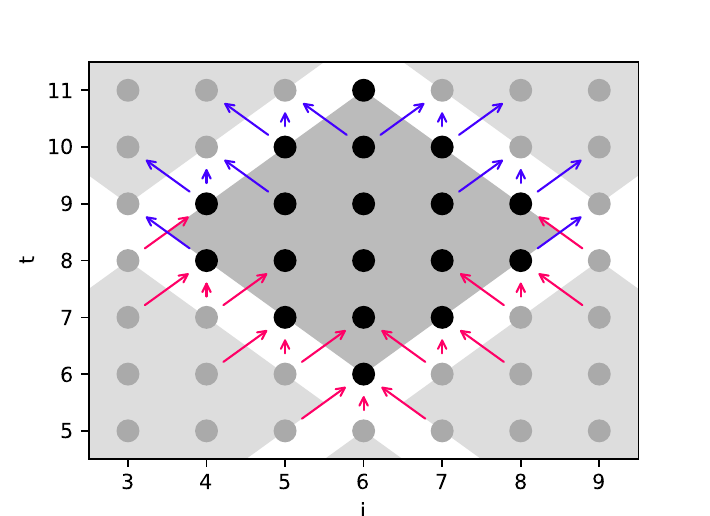}
    \caption{Inter-tile communication pattern for the Jacobi stencil: red arrows
    indicate data input into the tile shown in the center, and blue arrows
    indicate data output from this tile.}
    \label{fig:jacobi1d-dataflow}
\end{figure}

%For an easy understanding of the Jacobi stencil example of 
%Section~\ref{sec:running-example},

For this tiling scheme, intermediate results to be retrieved come from the tiles 
located below the tile to execute; in Figure~\ref{fig:jacobi1d-dataflow}, 
these tiles are designated as the source of incoming arrows into the 
tile to execute. Likewise, the outgoing arrows show those intermediate 
results that will be used by other neighboring tiles. All of these 
data transfers are the ones this work seeks to optimize; improvements of the
compute engine fall out of the scope of this paper.

\subsection{Deriving parallel accelerators using HLS}

The optimizations mentioned section~\ref{sec:locality-opt} are only a part
of all those optimizations that need to be applied to get the best
performance. One also needs to extract parallelism to maximize utilization
of operators on the FPGA, and create a macro-pipeline to maximize the 
compute throughput.

\begin{figure}
    \centering
    \includegraphics[width=0.8\columnwidth]{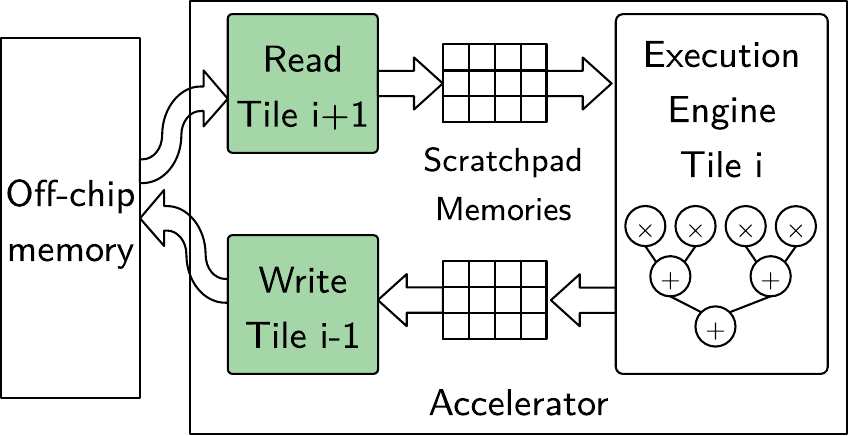}
    \caption{Macro-pipeline structure: read-execute-write. Our contribution
    focuses on the read and write stages.}
    \label{fig:macro-pipeline}
\end{figure}

Loop tiling naturally yields a ``read-execute-writeback'' 
macro-pipeline structure as illustrated in Figure~\ref{fig:macro-pipeline}: because tiles can be executed atomically, all 
I/O operations can happen before and after execution. HLS tools such as 
Vitis HLS support such macro-pipelines through manual code annotation, but through a restricted set of conditions of the pipeline (i.e. absence of cyclic dependency between stages). Moreover, automated macro-pipelining further increase pressure on the on-chip memory usage, as buffers used for inter-stage communication are either implemented using FIFOs or duplicated.
While standard coding techniques would pass the complete tile data buffer across pipeline stages, this widely inefficient in hardware as communication only requires a subset of the actual tile data.
On one tile of our Jacobi-1D example from subsection~\ref{sec:running-example}, the required data to be communicated is represented in blue in Figure~\ref{fig:jacobi1d-mars}.

Usually, the goal of the execute stage is to take advantage of massive operation-level parallelism (thanks to loop pipelining and unrolling), that have already been
extensively addressed~\cite{Pouchet_2013, sohrabizadeh2022autodse, choi2018hls, santos2020automatic}.
In this paper, we only seek to optimize transfer times and memory bandwidth usage; optimisation of the complete design including crafting and balancing of a coarse-grain pipeline are not evaluated. 

In applications with low operational intensity, the limited off-chip memory bandwidth turns the read and write stages into performance bottlenecks, even with aggressive tiling transformations. In most cases, this is due to a poor utilization of the off-chip memory interface, where only a fraction of the peak bandwidth is effectively used due to inefficient access patterns. 
As a matter of fact, approaching the peak memory bandwidth requires that almost all access to external memory consist of large transfers over contiguous memory locations (called memory \emph{burst}).

HLS tools can infer burst memory accesses depending on the target interface. In the case of a shared bus (e.g. AXI, PCIe), which is commonly found for off-chip accesses, a \emph{burst} access may occur if the bus supports it and the compiler recognizes access to a series of consecutive addresses. 
Tools such as Vitis HLS 2022.2 exploit this using with either a call to a HLS-specific \texttt{memcpy} routine, or through some form or pattern matching in the source code.
In burst mode, no cycle is spent stalling for a new value after a one-shot initialization latency, which yields full utilization of the available bandwidth.

The goal of this work is to propose a \textbf{source level compiler optimisation to (i) reorganize data in memory to enable contiguous burst access and (ii) further improve bandwidth utilization through packing and compression.} Our optimization pass is meant to be integrated within an HLS polyhedral compilation flow, as illustrated in Figure~\ref{fig:polyhedral-flow}; aiming at sitting between the locality optimization phase (tiling) and the HLS synthesis stage. In fact, \textbf{our approach does not replace locality optimizations}, it \textbf{complements them}.
%$$ \boxed{
%\begin{array}{c}
%\textbf{Our approach does not replace locality optimizations,} \\
%\textbf{it is complementary to them.}
%\end{array}
%} $$

\begin{figure}
\centering
\includegraphics[width=\columnwidth]{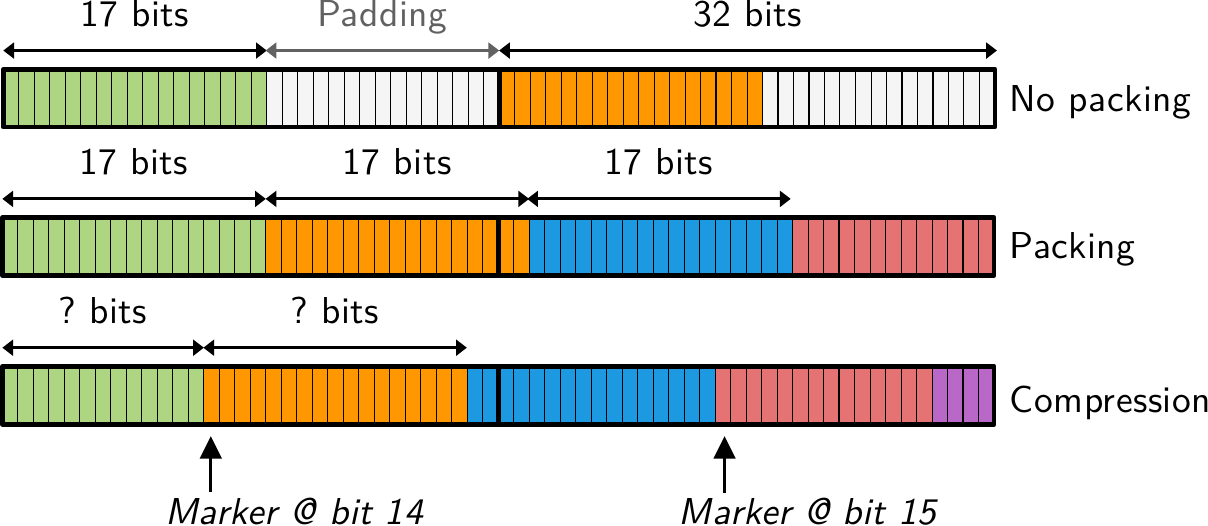}
\caption{Data packing and compression reduce storage and transfer redundancy 
at the expense of address alignment and, for compression, predictability of addresses.}
\label{fig:packing-compression}
\end{figure}

\subsection{Padding vs packing} 

In order to maximize the utilization of bandwidth, every bit of data transmitted must be useful. However, with  domain-specific data types (e.g., custom fixed point), unused bits must usually be transmitted due to memory alignment requirements. In the following, we explain how data contiguity can be leveraged to two ways: packing data to reduce the unused bits transmitted; and compressing
data to further save bandwidth.

Most memories are byte-addressable and most processor architectures also require 
aligned accesses at word boundaries, usually at 32 or 64 bits.
Although FPGA accelerators can operate on arbitrary-precision data types, off-chip
data transfers must abide by the addressing requirements of the external memory.
They therefore need to pad the incoming and outgoing data: in practice, for a 17-bit access, 32 bits of data will be transferred, 15 bits of them being wasted in padding.

Note that padding is necessary to enable random accesses to data: it provides the
guarantee that a given memory cell contains only the requested data and no
manipulation needs to be done to extract it. Data accessed in a contiguous manner
does not need this guarantee and may overlap multiple adjacent cells, as simple
wire manipulations on the FPGA will give back the original data.

Data packing, as illustrated in Figure~\ref{fig:packing-compression},
consists in avoiding padding the data so that words are adjacent at the bit
level in memory.
Figure~\ref{fig:packing-compression} shows buffer structure for unpacked and packed data of
17 bits in 32-bit words. Unpacked data has aligned addresses, but
requires extra storage and transfers unused data; packed data has 
unaligned addresses but saves storage and avoids some redundant bits from being transmitted. It becomes however impossible to randomly seek in a packed stream due to misalignment without additional data processing, but by definition such random seeks do not happen with contiguous accesses.

In our approach, we leverage contiguous accesses to (i) avoid the adverse effects of packing induced misalignment and (ii) to maximize bandwidth utilization by not padding data.

\subsection{Runtime data compression}
\label{sec:compression-alg}

Packing data saves bandwidth by eliminating the padding bits, and
is applied independently of the data itself. However, further optimisation
is possible by exploiting properties of the data (e.g., correlation between
integers in an array) for compression.
When this technique is applied right before / after off-chip communications,
the design benefits from a reduction of I/O cycles
(as the amount of data transferred is reduced) without increase of the computation
subsystem as the latency of the compression
module can be hidden by the pipelining structure

%It is possible in FPGA accelerators to compress blocks of data in
%a pipelined fashion before storing them into off-chip memory, 
%thereby saving I/O cycles, and decompress the data when it 
%comes back on chip. Thanks to pipelining, the latency of runtime 
%compression is hidden.

Compression is easy to apply to contiguous streams of data, but is
not to data where indexed or random accesses are necessary. 
We must exhibit access contiguity, as it is in general impossible 
to seek within a compressed block without decompressing more data 
than needed. Figure~\ref{fig:packing-compression} shows that the
position of data within a compressed block is unpredictable.

Our approach performs runtime compression and, to maximize its 
efficiency, creates data blocks with a contiguous access guarantee
to ensure every decompressed piece of data is used.

% \begin{itemize}
% \item exploiter la correlation des données (par ex entiers)
% \item compression sur un flux continu OK, mais impossible sur un tableau (car besoin d; indexation).
% \item Mais on peut compresser au niveau block en off-chip memory, et decompresser le block quand ecrit en memoire on chip (et vice-versa)
% \end{itemize}
% It is possible to further save data transfers by compressing data within a  contiguously accessed block. 

% However, after block-level compression, not only are addresses unaligned, but they also become unpredictable, due to variable length size coding. In general  random seeks into a compressed data stream is impossible, unless there are markers (or a similar mechanism) to keep track of specific addresses within the stream.

% Figure~\ref{fig:packing-compression} illustrates the bandwidth savings and the need for either contiguity or markers: otherwise it is impossible to find the  position of a word inside of a compressed stream of values.

% Our algorithm exploits both compression and packing. It creates data blocks with guaranteed contiguous accesses, compresses them and packs the compressed blocks together, provide enhanced bandwidth savings.

%\subsubsection{Compression algorithm}

In general, the compression algorithm is domain-specific, 
e.g., ADPCM for voice~\cite{Cummiskey_1973} or JPEG for images~\cite{Skodras_2001}. 
For FPGA implementations, the choice of the algorithm is also 
driven by its throughput: compression and decompression 
must be able to sustain the input and output throughput not to
become the bottleneck.
%Our approach is agnostic to this\Todo{Pas d'accord, l'algo doit pouvoir compresser/décompresser à la volée donc efficace} and 
We choose to illustrate the idea with a simple differential 
compression algorithm which encodes a sequence 
$w_0 w_1 \dots w_n$ of $N$-bit words as follows:

\begin{itemize}
    \item Encode $w_0$ as is.
    \item For $1 \leqslant i \leqslant n$:
    \begin{enumerate}
        \item Compute $\Delta = w_i - w_{i-1}$,
        \item Let $L$ be the number of leading zeroes of $\Delta$ if $\Delta \geqslant 0$,
        or leading ones if $\Delta < 0$,
        \item Encode $N-L$ using $\lfloor1+\log_2(N)\rfloor$ bits, followed by the sign bit of $\Delta$,
        \item Encode the $N-(L+1)$ lowest bits of $\Delta$.
    \end{enumerate}
\end{itemize}

This technique is especially effective when the distribution of the transferred data is not spread, typically on benchmarks based on the computation of the average such as our Jacobi-1D example from subsection~\ref{sec:running-example}.

\section{Memory Layout Optimization}
\label{sec:layout}

%\Todo{It is important to retain clarity for double blindness: THEY did MARS, and WE are doing this that and the other}

% \Todo{Pour moi, ce qui suit, c'est plutôt la section 4 ça non ?, il faut plustôt expliquer que le "cornerstone" de l'approche c'est un layout mémoire qui maximise la contiguité et minimise le nombre de blocks}
% CF : je pense que c'est corrigé

This work seeks to minimize the I/O cycles for an accelerator to transmit 
and retrieve its data into and from global memory. Locality optimizations
having already been applied, we do not want to store or retrieve 
\emph{fewer} values from memory, but rather make \emph{better} accesses
to memory, from the bandwidth utilization, and therefore I/O cycles, standpoints.

To this aim, we build \textbf{a contiguous, irredundant and compressed data layout}.
This section details the steps taken: first, we analyze the program
and extract sets of on-chip data as contiguous data blocks; second, we lay
out these data blocks to obtain further contiguous accesses; third, we 
compress and pack the data blocks together to save even more bandwidth.

\subsection{Extracting Contiguous Data Blocks}
\label{sec:mars-extraction}

\begin{figure}
    \centering
    \includegraphics[width=\columnwidth]{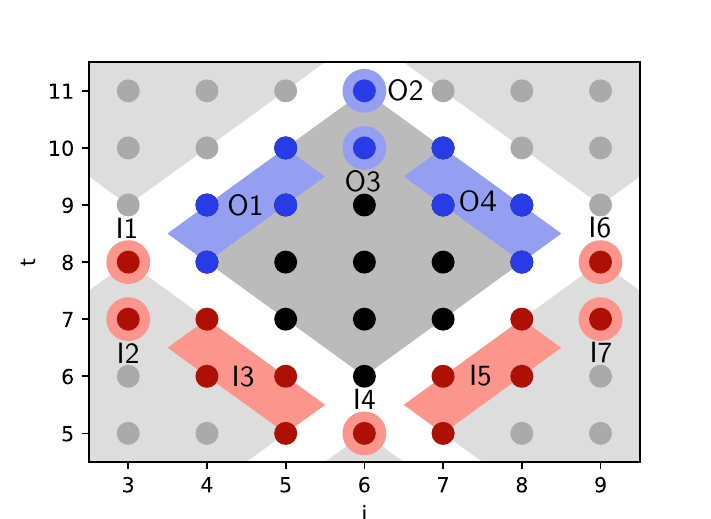}
    \caption{MARS: Groups of points within a tile which data is contiguous
    in global memory. In blue, the MARS produced by the center tile (\textsf{O1}
    to \textsf{O4});
    in red, the MARS consumed by that same tile (\textsf{I1} to \textsf{I7}).}
    \label{fig:jacobi1d-mars}
\end{figure}

The first step in our method consists in analyzing a program's behavior
with respect to memory, to determine which data can/should be grouped
together as contiguous blocks. The sought groups of data honor two
properties:

\begin{itemize}
    \item Atomicity: If any data in the group is needed for an instance
    of the accelerator's execution flow (a tile), then the entire group
    is also needed for the same tile.
    \item Irredundancy: No data is retrieved or stored more than once 
    into memory throughout the execution of a single tile.
\end{itemize}

These groups of data are determined by using the analysis technique from 
Ferry et al. \cite{Ferry_2023} within a polyhedral compiler.
This analysis yields sets of on-chip memory addresses, such that all 
the data from these on-chip cells will be allocated a contiguous block
of data in off-chip memory.

\subsubsection*{Example}
\label{sec:running-example-mars}
Applying the MARS analysis from \cite{Ferry_2023} to the Jacobi stencil
of Section~\ref{sec:running-example} gives the sets of addresses 
corresponding to the points illustrated in Figure~\ref{fig:jacobi1d-mars}:
\begin{itemize}
    \item For the input of each tile, seven contiguous blocks of data
    labeled \textsf{I1} to \textsf{I7} are to be taken, across three different producer tiles.
    \item For the output, each tile will produce four contiguous blocks
    of data labeled \textsf{O1} to \textsf{O4}.
\end{itemize}
There is a correspondence between output blocks (MARS) O$x$ from a tile and input
blocks I$y$ from other tiles: each O$x$ corresponds to one I$y$ in several
other tiles.

Without any further information, the result of MARS analysis would make
the accelerator require seven input and four output burst accesses. This
number could potentially be reduced. If \textsf{I1}, \textsf{I2} and \textsf{I3}
were adjacent in memory,
it would be possible to make a single access instead of three, and likewise
for \textsf{I5}, \textsf{I6} and \textsf{I7}. The total number of input accesses would go down to 
just three.

In order to reduce the number of accesses to the above, we have to show
that it is actually achievable: the blocks \textsf{I1} through \textsf{I7} are read by
multiple tiles, and coalescing opportunities for one tile may be 
incompatible with another tile's coalescing opportunities.

The next subsection formalizes this example into an optimization problem
seeking to minimize the number of accesses.

\subsection{Enabling Coalesced Accesses across Contiguous Data Blocks}
\label{sec:lp}

From the polyhedral analysis of the previous subsection, we have
determined sets of on-chip data to be grouped as contiguous blocks
of data, called MARS. How these blocks are laid out in memory 
is important for access performance: if multiple MARS happen to be
accessed in a row and they are adjacent in memory, the accesses to
these MARS can be coalesced into a single access and better utilize
bandwidth.

This section explains how the ``outer layout'' of the MARS is 
determined so as to maximize the coalescing opportunities.

\subsubsection{Properties of the layout}
\label{sec:lp-hypotheses}

The goal of this work is to minimize the number of I/O cycles, and 
therefore the data layout must exhibit contiguity (for both reading
and writing). However, that contiguity must not come at the price of
an increase in I/O volume. To model this constraint, we apply two
hypotheses.

\paragraph{Contiguous tile-level allocation}
\label{sec:hyp-contiguous-allocation}
We are looking for a layout of MARS in memory, and know that 
compression will be applied to them. Due to the size and position of 
compressed blocks being unpredictable, it is not feasible to
interleave MARS from multiple tiles in memory. Therefore,
\textbf{we allocate each tile a contiguous block of memory for 
its MARS output}.

This allocation has two consequences: the write side can be done entirely
contiguously, and we only have to optimize contiguity at the read side.

\paragraph{Irredundancy of storage}
\label{sec:hyp-irredundant-storage}
Under the previous hypothesis, we want to maximize the coalescing 
opportunities between MARS accesses for the read side only.
While it is possible to obtain this contiguity by replicating the
MARS in multiple layouts, one per consumer, doing so would defeat 
the goal to save I/O cycles.
We therefore choose to \textbf{store each MARS only once in memory} 
(irredundant storage).

The goal is now to find a single layout for the MARS produced by each
tile, that exhibits as much read-side coalescing opportunities
as possible. We obtain it throgh an optimization problem that is
defined in the next subsections.

\subsubsection{Example}
\label{sec:running-example-lp}

In the example of Section~\ref{sec:running-example-mars}, it appeared that the number
of burst accesses could go from 7 to 3. Let us show there actually exists
a layout achieving these 3 bursts.

Figure~\ref{fig:jacobi1d-mars} shows the correspondence between input
and output MARS:
\begin{itemize}
    \item \textsf{I1}, \textsf{I2} and \textsf{I3} come from the southwest tile, corresponding to its \textsf{O2}, \textsf{O3} and \textsf{O4} blocks. We would like these three MARS to be contiguous, regardless of which relative order, to make a single burst.
    \item \textsf{I4} comes from the south tile, corresponding to its \textsf{O2} block.
    \item \textsf{I5}, \textsf{I6} and \textsf{I7} come from the southeast tile, corresponding to its \textsf{O1}, \textsf{O2} and \textsf{O3} blocks. We would also like them to be contiguous.
\end{itemize}

We do not make any hypothesis on the relative location of data from the
southwest tile, the south tile and the southeast tile. This makes it 
impossible to obtain fewer than 3 burst accesses.

% \Todo{PB d'articulation entre les deuc section. La notion de "solver" sort d'un peu nul part, il faudrait remettre en contexte}

The information we have at this point can be used as the constraints 
and objective of an optimization problem: we want to maximize the
number of contiguities in the layout among those desired, under the
irredundancy constraint. We provide a solver with the following problem:
\begin{itemize}
    \item Maximize the contiguities among the desired ones: make MARS \textsf{O2}, \textsf{O3} and \textsf{O4} contiguous in any order, and make MARS \textsf{O1}, \textsf{O2} and \textsf{O3} also contiguous in any order.
    \item Per the hypothesis of Section~\ref{sec:lp-hypotheses}, we want a layout of MARS \textsf{O1}, \textsf{O2}, \textsf{O3} and \textsf{O4}.
    \item There can be no fewer read bursts than 3.
\end{itemize}

The solver returns the following layout of the output MARS for each tile: \textsf{O1}, \textsf{O3}, \textsf{O2}, \textsf{O4}.

Looking from the consumers,
\textsf{I1}, \textsf{I2} and \textsf{I3} (resp. southwest \textsf{O2}, \textsf{O3} and \textsf{O4}) are 
contiguous; \textsf{I5}, \textsf{I6} and \textsf{I7}
(southeast \textsf{O1}, \textsf{O2} and \textsf{O3}) are also contiguous.
We can therefore coalesce, 
for each tile, the reads \textsf{I1}, \textsf{I2} and \textsf{I3}
into a single burst, and \textsf{I5}, \textsf{I6} and \textsf{I7} into another
burst, achieving the three sought input bursts.

\subsubsection{General case}
%In order to get coalesced MARS accesses, we have to lay out the MARS in the global memory. Every MARS is guaranteed to be transferred atomically as a burst, yet the input of each tile is composed of several MARS.

In this section, we lay out the blocks of data from the MARS analysis to
maximize the coalescing opportunities between them.

With the allocation choice of Sec.~\ref{sec:lp-hypotheses}, writes are
guaranteed to be done without discontiguity. We therefore lay out the MARS
to make the \emph{read} side as contiguous as possible. In other words, we
need to lay out the MARS  in memory so that as many MARS as possible can be
read as a \emph{coalesced} burst.

We propose to model this problem as an Integer
Linear Programming optimization problem as described in Algorithm~\ref{alg:lp}.
Intuitively, if a pair of MARS is needed by a consumer tile and the two MARS
are next to each other in memory, then a coalesced access for the two 
(a ``contiguity'') is issued. We therefore seek to maximize the number of such contiguities.

\begin{algorithm}
\small
\SetAlgoLined
\KwIn{$M_I = (m_i) : i=1,...,N$ = list of MARS,\\
$\mathcal{P}$ = list of producer tiles from which MARS are needed,\\
}
\KwResult{$M_O$ = ordered list of MARS}

\textbf{Optimization Variables}:\\
\textbf{let} $\delta_{i, j} \in \lbrace 0,1 \rbrace$ = successor variables: $\delta_{i,j} = 1$ encodes that MARS $i$ is immediately before MARS $j$ in memory. \\
\textbf{let} $\gamma_{i} \in \lbrace 1, \dots, N \rbrace$ be a permutation; $\gamma_i$ is the position where MARS $i$ will be in the final layout.

\textbf{Problem Constants:}\\
\textbf{let} $a_{p,i,j}$ be equal to 1 if $m_i$ and $m_j$ ($i,j \in [\![1, N]\!]$) from tile $p \in \mathcal{P}$ are consumed together, 0 otherwise. \\

\textbf{Maximize} \#\{contiguities\}: $\sum_{P \in \mathcal{P}} \sum_{i=1}^{N} \sum_{\substack{j=1\\j \neq i}}^{N} a_{P,i,j} \delta_{i,j}$ 
\textbf{subject to:}
\begin{itemize}
\item $\forall i : \delta_{i,i} = 0$ (a MARS is not its own predecessor)
\item $\forall i : \sum_j \delta_{i,j} \leqslant 1$ (a MARS has at most 1 precedecessor)
\item $\forall j : \sum_i \delta_{i,j} \leqslant 1$ (a MARS has at most 1 successor)
\item $\sum_i \sum_j \delta_{i,j} = N-1$ (number of successor relations)
\item $\forall i : 0 \leqslant \gamma_i \leqslant N-1$ (permutation of length $N$)
\item $\forall i, j : (\gamma_j - \gamma_i = 1) \Leftrightarrow (\delta_{i, j} = 1)$ (definition of successor)
\item $\forall i, j : (i \neq j) \Rightarrow |\gamma_i - \gamma_j| \geqslant 1$ (MARS have $\neq$ positions)
\end{itemize}

\Return{$M_O = (m_{\gamma(1)}, ..., m_{\gamma(N)})$ ($\gamma$-ordered list of MARS)}
\caption{Optimizing the MARS layout}
\label{alg:lp}
\end{algorithm}

The solution to this optimization problem, given by the solver 
%Gurobi solver\Todo{Le fait que ce soit Gurobi ne change rien à la réponse. La mention à Gurobi doit-être faite dans la partie exp. results plutôt qu'ici},
is an ordered list of the MARS produced by each tile, that allows the 
minimal number of transactions to read all MARS input of a tile.

The layout created in this section honors the \emph{irredundancy} 
property of the MARS (see Section~\ref{sec:mars-extraction}), but 
does not yet take full advantage of their \emph{atomicity}: 
the fact that the MARS are contiguous blocks of data makes them
ideal candidates for data packing and compression. This is what 
we perform in the next subsection.

\subsection{Contiguity-Preserving Block Compression}

\begin{figure}
\centering
\includegraphics[width=\columnwidth]{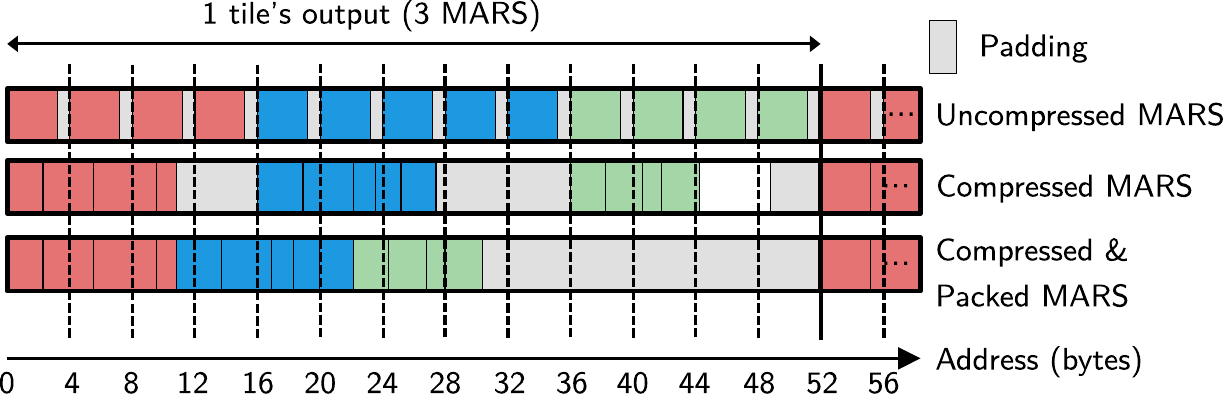}
\caption{MARS data shown without compression, with compression 
(inside the MARS) and with MARS compression and packing. Packing 
the compressed MARS preserves the contiguity of coalesced accesses.}
\label{fig:mars-compression-packing}
\end{figure}

So far, our approach has given a layout of data in memory
enabling coalesced accesses to contiguous blocks of data produced
and consumed by an accelerator. These blocks have an atomicity
property that we can further exploit to save bandwidth, by
applying data packing and compression, as illustrated in 
Figure~\ref{fig:mars-compression-packing}.

\subsubsection{Combining compression and packing}

Compressed blocks of data must be considered atomic in the sense that no
random accesses into them are possible. This atomicity property is borne
by the MARS, as each MARS data block is entirely used when it is
accessed, i.e. there are no partial accesses to a MARS.

Compressing the MARS reduces the size of the data and therefore saves
bandwidth and storage space; however, it can also break the contiguity 
brought by the layout of Section~\ref{sec:lp} as illustrated by Figure~\ref{fig:mars-compression-packing}. To preserve it,
we also apply \emph{packing} to the compressed MARS, making them 
immediately adjacent to each other in memory. Packing compressed MARS 
also spares the accelerators from unused reads due to padding.

\subsubsection{Need to preserve metadata}

As the size of compressed blocks depends on their data, it is impossible
to know the exact size of each access. However, the size of a burst access
must be known prior to the request being issued; additionally, using an 
estimation of the size or an over-approximation would result in unused 
input data or additional requests to fill in missing data. 

In order to be able to exactly fetch the right size, it is necessary to
keep track of the size of each compressed MARS.
Moreover, the packing of compressed MARS means that the start of a 
compressed block may be improperly aligned. It is also therefore necessary
to keep track of the alignment of each MARS for proper decompression. 
In our implementation, bookkeeping is done using on-chip \emph{markers}
that are filled in after each MARS is compressed. 
Details are in Section~\ref{sec:compression-marker-impl}.

Packing will cause unused input data to enter; however, its size is bounded to
one aligned word at the beginning and one aligned word at the end of each
transaction. This input redundancy is notably independent of the size of
the MARS.

\section{Integration into HLS design flow}
\label{sec:implementation}

In this section, we show how we transform an HLS accelerator description
in order to optimize its off-chip memory accesses for bandwidth utilization. 

The off-chip data layout and compression proposed in Section~\ref{sec:layout} 
can be automatically implemented around the existing description of a tile in HLS.
The result is a sequence of steps:
\begin{itemize}
    \item Read MARS layout data and non-MARS input data from off-chip memory into on-chip FIFOs,
    \item Decompress the input data into FIFOs,
    \item Dispatch MARS data into on-chip buffers with an allocation suitable for computation, 
    \item Perform the computations onto on-chip buffers,
    \item Collect MARS output data from the on-chip buffers into FIFOs, 
    \item Compress the collected data,
    \item Write back the results into MARS layout in off-chip memory.
\end{itemize}

The next subsections explain how the complex data structures describing the
MARS are turned into two simple decompression/dispatch and collect/compression
steps.

%Implementing a MARS-enabled accelerator is done twofold: we need to implement both
%the on-chip computation and memory allocation, and input/output functions for MARS.
%In this section, we will focus on the I/O optimizations and how we can obtain burst
%accesses from polyhedral data structures as complicated as the MARS.

%\subsection{Polyhedral Optimizations for FPGA implementation}
%
%The compute and I/O functions of MARS-enabled accelerators are generated by 
%polyhedral code generators. However, polyhedral code generators \cite{Bastoul_2004} 
%usually produce complex control flows that depend on the tile's position within
%the iteration space. This subsection explains what simplifications we made and
%their consequences.
%
%\subsubsection{Common computational core}
%\label{sec:execution-core}
%The computational core of a MARS-enabled accelerator (a tile) must have a simple control
%flow to expose parallelism. 
%
%Thanks to the uniform dependence hypothesis, we make most tiles' control flow
%independent of the tile's position in the iteration space, except those at space
%boundaries. This results in a single loop nest with a simple control flow,
%which is sufficient to describe all non-boundary tiles.

%These are inferred from either calls to \texttt{memcpy},
%or from loops that increase a pointer with a unit stride.

\subsection{From MARS to Collect/Dispatch Functions}
The input and output data of each tile is respectively copied 
into and out of on-chip buffers before the tile execution takes place 
and after it has fully completed. This is the step where the data
goes from a contiguous layout to a non-contiguous layout (suitable
for execution) and vice-versa. 

Implementing these dispatch and collect steps requires to describe
each MARS so that the data contained in it is placed into, or taken
from, the right location in on-chip memory. Before dispatch and 
after collect, the data is located into FIFOs in the contiguous 
layout.

MARS can have arbitrary complex shapes, and cannot in general be 
described using simple loops. However, it is possible to fully unroll
these loops and obtain a list of on-chip addresses for each MARS. 
Such unrolled lists are placed into read-only memories on chip.
Iterating through these ROMs as in Figure~\ref{fig:mars-input-code}
gives the corresponding addresses. The size of these ROMs is notably
only dependent on the tile size, and not on the problem size or data
type.

%  \Todo{Est-ce que la taille de cette mémoire croite avec la taille du domaine d'itération, ou bien il les info peuvent être trasnlatée pour chaque tuile ?}

%For burst accesses to be extracted, the control flow of transfers has to be a single, perfect loop nest without guards. However, 

%The complexity of MARS only affects on-chip accesses and not off-chip accesses, where
%each MARS is a contiguous block of data. The solution to get bursts is to hide the complexity of the MARS loop nests, and use MARS size to perform contiguous accesses. 

%The tile size is a compile-time parameter; therefore, the position of each MARS' data within on-chip buffers is known at compile time. We therefore statically compute the on-chip position of all MARS data and place it in a ROM array. 

%MARS input and output operations are performed by constant-size loop nests as illustrated in Figure~\ref{fig:mars-input-code}, which take on-chip addresses from the ROM.

\begin{figure}
\begin{lstlisting}[style=CStyle, belowskip=-2 \baselineskip]
// MARS Dispatch
for(int i=0; i<33; ++i) {
  // take on-chip address from ROM
  struct mars_transfert mt = FPGA_MARS_IN_TBL[i];
  switch (mt.array) {
    // on-chip random write
    case MARS_DATA_ENUM::A: {
      marsToMem_A(mt.dim0, mt.dim1);
      break;
    }
    case MARS_DATA_ENUM::B: {
      marsToMem_B(mt.dim0, mt.dim1);
      break;
    }
  }
} 
\end{lstlisting}
\caption{Structure of the MARS dispatch implementation (off-chip to on-chip layout)}
\label{fig:mars-input-code}
\end{figure}

\subsection{Automatic compression}

When the data is in the contiguous layout in the form of MARS, it
can be seamlessly compressed and decompressed, and the compressed
MARS can be packed to preserve contiguity. We explain here the
compression, packing and decompression steps, along with how the
compression metadata is taken care of.

% Packing the MARS
% incidentally simplifies the compression and decompression but
% requires metadata to keep track of the location and size of 
% packed MARS.
%  \Todo{as discussde in 3.3.2. Au passage, tu ne dis toujours pas comment tu gères le pb. Ça va bcp frustrer le reviewer }

\subsubsection{Compressing Data and Packing MARS}

The compression step is relatively straightforward: the compression module
takes its input from the collect step FIFO, and generates a compressed 
stream of data from it. The layout of the data in this FIFO is not altered
by the compression step. 
Likewise, the decompression step takes a stream of compressed words and
decompresses it into a FIFO, which is then used by the MARS dispatch step.
MARS packing is transparently implemented by the compression step: because
MARS are provided in a contiguous manner from the collect step, the first
word of each MARS will be immediately adjacent to the last word of the 
previous MARS in the compressed data stream.

Our compressor, which algorithm is given in Section~\ref{sec:compression-alg},
is pipelined with an initiation interval of 1 cycle, despite a loop-carried
dependence. 
%\Todo{Dire que la compression/décompression se fait facilement avec un II=1, malgré la dépendance circulaire (au pire on peut compresser la différence  entre x(n) et x(n-2) }

The difficult part to implement is decompression: because not all MARS from 
a given tile are decompressed, we need to be able to seek at the start of a 
particular MARS. This ability is given by metadata described in the next
paragraph.

% Markers are therefore stored on chip to this aim; the next paragraph 
% explains how these markers are implemented. Using them, it is possible to
% seek within the compressed data, and therefore not start the decompression
% from the start.

\subsubsection{Metadata management}
\label{sec:compression-marker-impl}

The consequence of MARS compression is that their size is unknown 
a priori.
% \todo{je ne susi pas sur que unpredicatble soit le bon terme. Je dirai plutôt "known at runtime "}.
To preserve the contiguity of the layout from Section~\ref{sec:lp}, we
must avoid padding the compressed MARS to preserve alignment,
Therefore, the compressed MARS are packed and immediately adjacent to 
each other in memory.

To keep track of the position of each MARS, we use a data structure with
two pieces of information: a \emph{coarse-grain} position indicating how
far (in aligned words) to seek, and a \emph{fine-grain} position marker 
that specifies which bit is the first of the said MARS.

%\Todo{Une figure illustrant le truc serait la bienvenue}
Because the length of MARS is known at compile time and constant across 
tiles, the position of the markers within the uncompressed stream is also
constant. Therefore, like the MARS descriptions, the positions of markers 
(i.e. start of each MARS) within the uncompressed stream are put into a ROM:
\begin{lstlisting}[style=CStyle, belowskip=-2 \baselineskip]
#define NB_MARKERS 3
#define MARKERS {62, 63, 64}
\end{lstlisting}

The markers for the compressed stream are maintained within an on-chip cache,
which size is specified at synthesis time via a macro:
\begin{lstlisting}[style=CStyle, belowskip=-2 \baselineskip]
struct compressed_marker<NB_MARS_POS_BITS, LOG_BUS_WIDTH> markers[COMPRESSION_METADATA_SIZE][NB_MARKERS];
\end{lstlisting}

The allocation within this cache is done from the host: registers are used to 
specify whether a tile's MARS are compressed, whether its dependences are, and
where the markers for its dependences are located. This location depends on
the size of the space; for the Jacobi stencil, the formula is:
\begin{lstlisting}[style=CStyle, belowskip=-2 \baselineskip]
unsigned compressionMetadataAllocation(
 int tsteps, int n, int M1, int M2, int k1, int k2) {
  return (k2) + M2DEC_FORMULA + M2 * ((k1 - 1) & 0x01);
}
\end{lstlisting}

It should be noted that the \texttt{markers} structure is persistent
between runs. It is updated by the MARS write step and used by the MARS
read step. This update prevents the current HLS tools from constructing
a macro-pipeline (e.g. using the \texttt{HLS DATAFLOW} pragma) unless 
the structure is in a separate module.

% \subsubsection{Restriction}
% There is no task-level pipeline (\texttt{HLS DATAFLOW} pragma) in our accelerators,
% due to an analysis limitation of the HLS tool: the marker allocation is assumed by
% the tool to create a back-edge (loop-carried dependence) in the dataflow graph. 
% The allocation being controlled by the host, we can ensure the markers used by a
% tile are not used until that tile's execution is complete.
% %  \Todo{On ne comprend pas le message de ce paragraphe, et tu t'y tire un missile  dans le pieds. Il faut que tu explique que tu as  *hacké* Vitis pour y arriver, mais que c'est un hack "legal"}

\subsection{Host/FPGA dispatching of tiles}
The FPGA accelerator must have a simple control structure to exhibit as much 
parallelism as possible. Therefore, only \emph{full tiles} are executed on FPGA. 
Full tiles also all have the same volume of I/O, regardless of their position
in the iteration space. 

Partial tiles, i.e. those that contain space boundaries, are run on the host CPU, 
using the original program's allocation. 
To permit this, data computed on FPGA is taken back from MARS into the original 
program's memory, and MARS are created back from partial tiles results. It can 
be demonstrated that no FPGA tiles need any missing MARS data from partial tiles, 
and therefore there is no issue in writing part of the MARS for these tiles.

The operations performed to execute a partial tile (on the host) are:
\begin{itemize}
    \item Read MARS from neighboring full tiles that were executed on FPGA, remap their data to its original location,
    \item Execute the tile's iterations using the original allocation,
    \item Write back MARS by copying data from the original allocation, skipping cells that would be in MARS yet have no producer iteration.
\end{itemize}

The control flow necessary for compression would significantly lengthen the execution
of host tiles. Therefore, only tiles which producers and consumers are all executed
on FPGA will use compression.

\section{Evaluation}
\label{sec:evaluation}

\begin{table*}
\centering
\begin{tabular}{ccccccc}
Benchmark & Tile Sizes & \#MARS In & \#MARS Out & Read bursts & Write bursts\\
\hline
\texttt{jacobi-1d} & $6\times 6$ $64\times 64$, $200\times 200$ & 7 & 4 & 3 & 1 \\
\texttt{jacobi-2d} & $4 \times 5 \times 7$, $10 \times 10 \times 10$ & 28 & 13 & 10 & 1 \\
\texttt{seidel-2d} & $4\times 10\times 10$ & 33 & 13 & 10 & 1
\end{tabular}
\caption{Characteristics of the selected benchmarks. The number of bursts per tile 
accounts for layout-induced access coalescing and is independent of tile and problem size.}
\label{tab:mars}
\end{table*}

We evaluate our approach with respect to the following questions:
\begin{itemize}
    \item \textbf{Compile-time performance:} How much time does it take to compute the MARS layout?
    \item \textbf{Design quality:} How does using MARS affect the FPGA accelerator's area consumption?
    \item \textbf{Runtime performance:} How much I/O cycles do compressed MARS save with respect to a non-MARS memory layout?
    \item \textbf{Applicability:} How does the data type, tile size and problem size affect the compression ratio?
\end{itemize}

\subsection{Protocol and benchmarks}

\subsubsection{Benchmarks}

We have selected the following applications from the PolyBench/C suite\cite{Polybench}:
\begin{itemize}
	\item \texttt{jacobi-1d}: Jacobi 1D stencil, as used in the running example;
	\item \texttt{jacobi-2d}: Two-dimensional version of the Jacobi stencil, exhibiting few and simple MARS;
	\item \texttt{seidel-2d}: More complex benchmark exhibiting a higher number of MARS with more complex shapes.
\end{itemize}

Layout determination was done using the Gurobi solver (version 10.0.3 build v10.0.3rc0 (linux64)).

The data types used are fixed-point numbers (18 bits, 24 bits, 28 bits)
and floating-point numbers (float, double). We also ran simulations with
a 12-bit fixed-point data type without synthesizing it, Vitis HLS being
unable to infer bursts from that data type.

The chosen applications provide a non-MARS data layout in their original code.
Because FPGA developers usually try to seek burst accesses where possible,
we have created two access patterns on the non-MARS layout to compare against
that try to exhibit bursts:
\begin{itemize}
    \item A minimal access pattern, fetching and storing the exact I/O footprint of the tile, letting the HLS tool infer bursts where possible.
    \item A rectangular bounding box of the accessed data like done in PolyOpt/HLS \cite{Pouchet_2013}, which description is simple enough to infer only burst accesses.
\end{itemize}
Both access patterns are generated using a polyhedral code generator
available in ISL \cite{Verdoolaege_2016}.

\subsubsection{Hardware platform}

We used a Xilinx ZCU104 evaluation board, equipped with a \texttt{xczu7ev} MPSoC.
We ran Pynq 3.0.1 with Linux 5.15 and synthesis was done using the Vitis/Vivado suite version 2022.2.2
%on Linux 3.10.0-1160.83.1.el7.x86\_64.
All benchmarks, are running at a clock frequency of 
187 MHz 
and
communicate with the off-chip DDR using one non cache-coherent AXI HP port.

\subsubsection{Protocol}

Each benchmark is run for each data type, each space size and tile size. 
Part of the computation is done on the host: incomplete tiles are executed on a single thread on the Cortex-A53 CPU of the MPSoC. Transfer cycles are measured only for the
FPGA tiles and do not account for the host.

Cycle measurements are gathered using an on-FPGA counter and the area measurements
are extracted from Vivado place and route reports.

Table \ref{tab:mars} shows the characteristics of each benchmark, in terms of number of MARS, and number of bursts after coalescing optimization of Sec.~\ref{sec:lp}.

\subsection{Results and discussion}

\subsubsection{Compile-time performance}

Table \ref{tab:compile-time-performance} shows the time it took for
each benchmark to be run through the layout determination and 
code generation framework.
The compilation process does not take more than a few seconds to 
execute for the benchmarks we selected, starting from the polyhedral
representation of the program to the end of HLS code generation.
Notably, the layout determination ILP problem only depends on the number of
MARS and is independent of the tile size.

\begin{table}
\centering
\begin{tabular}{ccc}
Benchmark & Tile Size & Compile Time (s) \\
\hline
\texttt{jacobi-1d} & $6\times 6$ & 0.76 \\
\texttt{jacobi-1d} & $64\times 64$ & 0.68 \\
\texttt{jacobi-1d} & $200\times 200$ & 1.02 \\
\texttt{jacobi-2d} & $4\times 5\times 7$ & 5.57 \\
\texttt{jacobi-2d} & $10\times 10\times 10$ & 5.09 \\
\texttt{seidel-2d} & $4\times 10\times 10$ & 3.21
\end{tabular}
\caption{Layout Computation and Code Generation Time}
\label{tab:compile-time-performance}
\end{table}

\subsubsection{Design quality}

\begin{figure*}
    \centering
    \begin{subfigure}{0.45\textwidth}
        \centering
        \includegraphics[width=\textwidth]{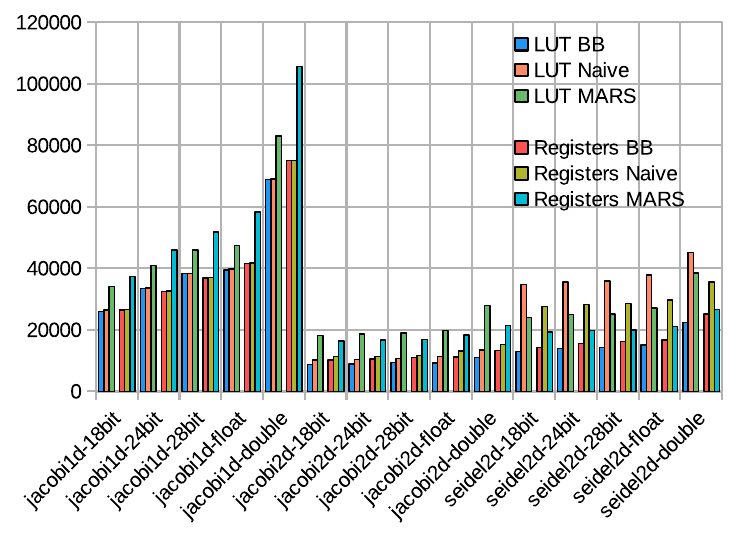}
        \caption{Logic resource occupancy: LUT, Registers}
        \label{fig:lut-reg}
    \end{subfigure}
    \begin{subfigure}{0.45\textwidth}
        \centering
        \includegraphics[width=\textwidth]{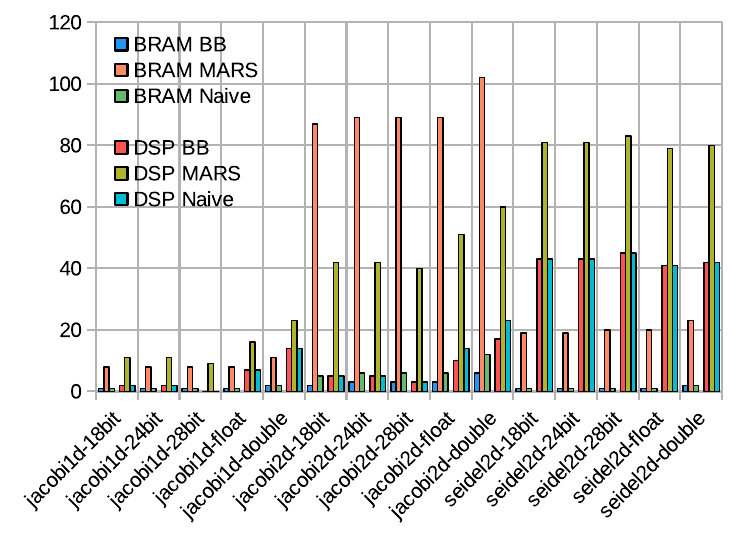}
        \caption{DSP and BRAM occupancy}
        \label{fig:dsp-bram}
    \end{subfigure}
    \caption{Area statistics for the benchmarks}
    \label{fig:area}
\end{figure*}

Figure~\ref{fig:area} shows the total area occupied by
our benchmarks, with respect to the different memory allocation
baselines. One tile size per benchmark is considered.

MARS introduces extra control logic and extra I/O 
functions that the other baselines do not have.
It is therefore normal to observe area increases
with this baseline. The most significant increases
in Figure~\ref{fig:lut-reg} are for \texttt{jacobi1d};
in this benchmark, on-chip arrays are implemented in
logic instead of Block RAM. Figure~\ref{fig:dsp-bram}
shows little DSP and BRAM consumption by this benchmark
compared to others. FIFOs holding all the MARS are 
implemented only on the MARS baseline, and require extra 
BRAMs. The extra DSP blocks for MARS baselines come 
from the address computations that are performed inside the 
I/O units; the size of the space is passed as a parameter 
instead of being a constant, requiring true multipliers.

Figure~\ref{fig:lut-reg} shows that the data width causes 
the logic area to increase with it. This increase is more
sensible in \texttt{jacobi1d} where the on-chip arrays are
implemented in logic instead of Block RAM, which effect is
also visible in Figure~\ref{fig:dsp-bram}.

% Table \ref{tab:area} shows a breakdown of the area for the
% \texttt{jacobi-2d} benchmark with respect to layout/access 
% pattern (naive, bounding box, MARS), and data type.

% \begin{table}
% \centering
% \begin{tabular}{cccccc}
% Layout & Data type & LUT & FF & BRAM & DSP \\
% \hline
% \texttt{naive} & 18-bit & 10265 & 11299 & 5 & 5 \\
% \texttt{naive} & 24-bit & 10345 & 11418 & 6 & 5 \\
% \texttt{naive} & 28-bit & 10746 & 11667 & 6 & 3 \\
% \texttt{naive} & \texttt{float} & 11359 & 13073 & 6 & 14 \\
% \texttt{naive} & \texttt{double} & 13411 & 15275 & 12 & 23 \\
% \hline
% \texttt{B-Box} & 18-bit & 8693 & 10181 & 2 & 5 \\
% \texttt{B-Box} & 24-bit & 8980 & 10572 & 3 & 5 \\
% \texttt{B-Box} & 28-bit & 9471 & 11010 & 3 & 3 \\
% \texttt{B-Box} & \texttt{float} & 9190 & 11246 & 3 & 10 \\
% \texttt{B-Box} & \texttt{double} & 11012 & 13343 & 6 & 17 \\
% \hline
% \texttt{MARS} & 18-bit & 18148 & 16437 & 87 & 42 \\
% \texttt{MARS} & 24-bit & 18589 & 16666 & 89 & 42 \\
% \texttt{MARS} & 28-bit & 19050 & 16877 & 89 & 40 \\
% \texttt{MARS} & \texttt{float} & 19774 & 18342 & 89 & 51 \\
% \texttt{MARS} & \texttt{double} & 27984 & 21508 & 102 & 60 \\
% \end{tabular}
% \caption{Occupied area for \texttt{jacobi-2d}, tile size 10x10x10.
% MARS versions contain a scratchpad memory for 11520 compression metadata.}
% \label{tab:area}
% \end{table}

\subsubsection{Runtime performance}

\begin{figure*}
\begin{subfigure}{0.45\textwidth}
    \centering
    \includegraphics[width=\textwidth]{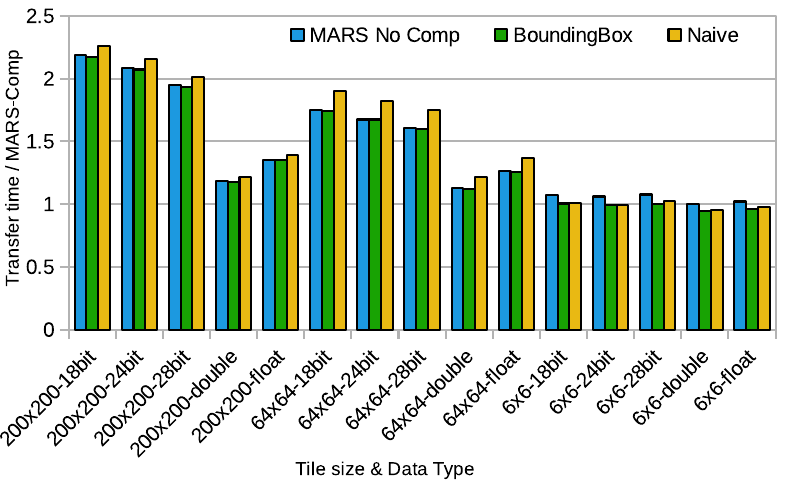}
    \caption{\texttt{jacobi-1d}}
\end{subfigure}
\begin{subfigure}{0.45\textwidth}
    \centering
    \includegraphics[width=\textwidth]{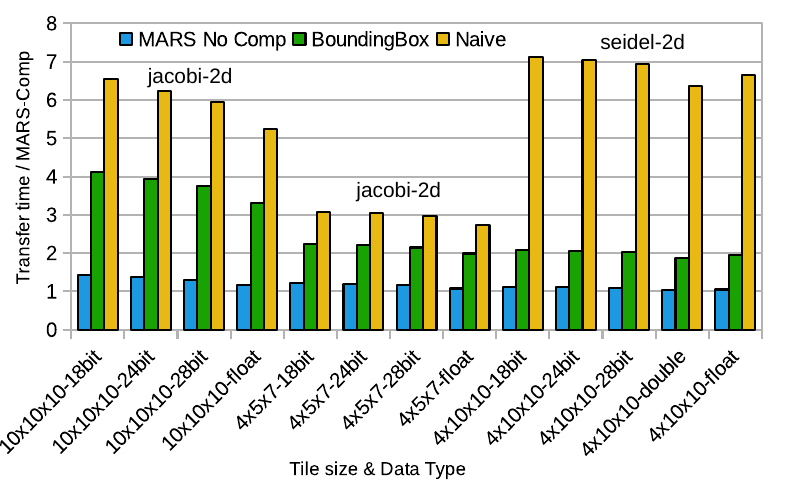}
    \caption{\texttt{seidel-2d}, \texttt{jacobi-2d}}
\end{subfigure}
\caption{Transfer time relative to compressed MARS (lower is better).}
\label{fig:transfer-time-vs-compressed}
\end{figure*}

Figure~\ref{fig:transfer-time-vs-compressed} shows the transfer time 
relative to compressed MARS for each data type and each benchmark. 

\paragraph{Impact of dimensionality}
For the \texttt{2d} examples that have 
three-dimensional iteration spaces, using MARS layout is already profitable 
versus the non-MARS layouts; most of the gains are due to contiguity more
than compression. On the one-dimensional Jacobi example, the gains are
on the contrary more due to compression: the data being one-dimensional,
non-MARS layouts are already contiguous. For small tile sizes like $6\times 6$,
the gains are marginal if any: the number of compressed elements is too
small to exhibit large gains from compression.

\paragraph{Effect of data type}
On the \texttt{jacobi-1d} benchmark, the choice of a $200\times 200$ tile size
shows a more significant benefit in using compressed MARS for fixed-point data
types than floating-point. This is explained with the better compression ratio:
when modeling continuous spaces like those used on the Jacobi stencils,
neighboring fixed-point values will have more higher bits in common than 
floating-point data where neighboring values mostly only share the exponent.

\subsubsection{Applicablity}

\begin{figure}
\centering
\includegraphics[width=\columnwidth]{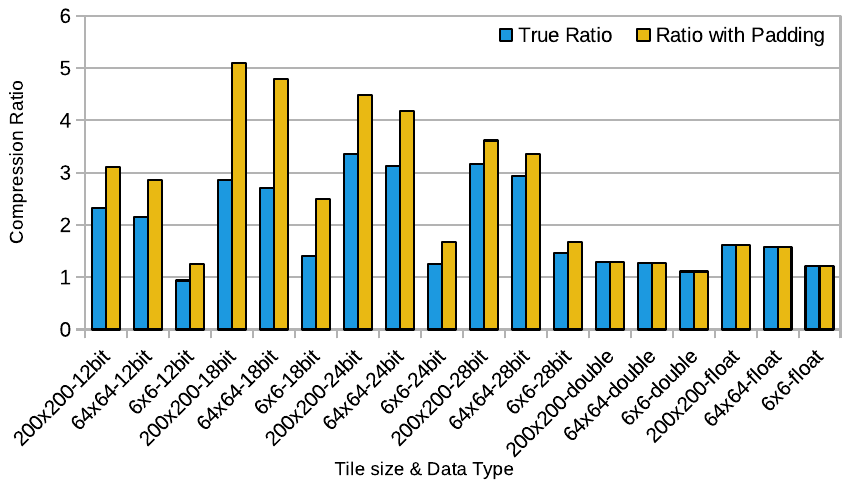}
\caption{Compression ratio vs. data type and tile size for jacobi1d}
\label{fig:compression-ratio}
\end{figure}

Figure~\ref{fig:compression-ratio} shows the compression rate
for each data type and tile size for the \texttt{jacobi1d} benchmark.
Two ratios are shown: the \emph{true ratio} which accounts only for
the bit savings due to compression, and a \emph{ratio with padding}
that accounts for the savings due to not padding the data.
The ratio with padding is the one our accelerators really benefit from,
because the data is not packed in memory except in compressed MARS form.

Overall, compressing the data for the selected benchmarks is almost
always profitable, possibly largely as the compression
ratio goes up to 5.09:1 for $200\times 200$ tiles and 18-bit type.

We can observe that large tiles ($64\times 64$, $200\times 200$) exhibit
closer compression ratios than smaller tiles ($6 \times 6$). This
discrepancy can be explained by the compressed chunks being too small to 
benefit from the data's low entropy; for the smallest data type and tile 
size, compressing data is even worse than not compressing. 

\section{Related Work}
\label{sec:related-work}

%\subsection{FPGA on-chip memory architecture}
%On-chip memory layout is addressed by a wide variety of works. On the one hand,
%the regularity of applications and their access patterns make templated memory 
%layouts a solution, presented in works such as Shafiq et al. \cite{Shafiq_2010}.
%On the other hand, automated application-specific array partitioning is available 
%in commercial HLS synthesizers such as Xilinx Vitis, and works such as Pouchet
%et al. \cite{Pouchet_2013} automatically explore various partitioning options
%to get the highest throughput.

%Co-optimization of on-chip and off-chip memory accesses can ensure there is no
%off-chip or on-chip memory-induced stalls. Sun et al. \cite{Sun_2022} take such
%an approach and design a deep neural network accelerator using an optimization 
%objective that maximizes on-chip port utilization while minimizing off-chip 
%access cycles.

This work comes as part of a global effort to relieve memory-boundness of
high-performance accelerators. In this section, we study other techniques
used to relieve the memory wall, some of which may not apply to compilers
due to not being automatable or breaking program semantics.

\subsection{Data Compression}
Data compression saves bandwidth without requiring to modify the program's
algorithm. It is therefore suitable for many bandwidth-bound problems.

\subsubsection{Compression techniques}

Data compression in FPGA accelerators is already a necessity for 
some intrinsically memory-bound applications such as deep convolutional 
networks, as no locality optimization can bring further bandwidth savings.
We here focus on two kinds of compression: lossless and lossy.

\paragraph{Lossless compression}
Lossless compression guarantees that the decompressed data is exactly the
same as the data before it was compressed. This property makes it possible
to do seamless, inline compression and decompression as is done for MARS. 
This is commonly performed in deep neural network accelerators \cite{Guan_2017,Aarrestad_2021}

Sparse encoding can be considered a form of lossless compression, and is
also commonly found in machine learning applications \cite{Du_2022,Li_2023}.
Sparse data structures often require indirections, which make them unsuitable
for use in polyhedral compiler flows unless the sparse structure is immutable
\cite{Horro_2022}.

\paragraph{Lossy compression}
It is possible to save more storage and bandwidth by using lossy compression.
Some applications in machine learning can afford a loss of precision without
degrading the quality of the result, e.g. using JPEG-compressed images 
\cite{Nakahara_2020} as inputs. However, automatic compression alters the data
and cannot be automatically inserted by a compiler unless the user explicitly 
requests it.

\subsubsection{Dynamic data compression}
In this work, we automate the compression and decompression of data
and it is transparent to the computation engine on FPGA. Other works \cite{Ozturk_2009,Sardashti_2016} perform dynamic, demand-driven 
compression without prior knowledge of the data to be handled.
Thanks to the static control flow of polyhedral codes, all the data 
flow is statically known and it is not necessary to maintain a cache 
policy.

\subsection{Memory access optimization}
The layout we propose in this work optimizes memory accesses by 
exhibiting contiguity using polyhedral analysis. In this section,
we go through other polyhedral memory access optimizations, and
explain other non-polyhedral ways it is possible to improve 
memory accesses.

\subsubsection{Polyhedral-based optimizations}
Using the polyhedral model and loop tiling to capture the data flow is the subject
of a number of works, proposing different breakups of the dataflow. Datharthri et al. 
\cite{Dathathri_2013} and Bondhugula \cite{Bondhugula_2013} propose decompositions
of the inter-tile communications to minimze MPI communications.
This work also seeks to optimize the passing of intermediate results, but the
data allocation is not statically determined like in this work.

A MARS-like decomposition of the inter-tile data flow into coarse-grain blocks for 
MPI has been proposed by Zhao et al. \cite{Zhao_2021}; our work achieves irredundancy
which requires a finer-grain modeling than the one proposed by \cite{Zhao_2021}.

\subsubsection{Domain-specific optimizations}
Memory access optimizations such as a change of data layout or access pattern can also be specific to each problem. We show here two cases of domain-specific optimizations.

\paragraph{Data blocking}
Data blocking (or tiling) is memory layout transformation that chunks multi-dimensional arrays into contiguous blocks. Similar to loop tiling, data blocking allows to coalesce 
accesses to entire regions of the input or output data. 

Data blocking can be efficient when the memory footprint of one iteration of 
an accelerator corresponds to a data tile. Although it has been used to 
optimize machine learning accelerators \cite{Tian_2020}, it may break spatial 
locality and degrade performance of accesses that cross tile boundaries. 

Data blocking can be combined with loop tiling and polyhedral analysis to coalesce 
inter-tile accesses. Ferry et al. \cite{Ferry_2022} seeks to exhibit the largest 
possible contiguous units spanning multiple tiles.

\paragraph{Stencil optimization}
Stencil computations have regular and statically known memory access patterns.
Domain-specific optimizers like SODA \cite{Chi_2018} derive an optimized 
FPGA architecture and memory layout specific to each stencil. 

% \subsubsection{Demand-driven access coalescing: scatter-gather units}

% Scatter-gather units are a way to obtain contiguity that does not 
% require compiler data flow analysis. Recent work on FPGA implementations
% of graph processing algorithms such as HitGraph \cite{Zhou_2019} and ThunderGP \cite{Chen_2021} particularly benefit from these units. 

% \subsubsection{Using caches}
% It is possible to use caches in FPGAs to handle irregular off-chip memory accesses 
% and avoid repeating off-chip accesses.
% For instance, Asiaciti and Ienne \cite{Asiatici_2019} propose a non-blocking cache 
% mechanism that simultaneously answers all outstanding requests to addresses within a
% cache line, while taking advantage of the distributed memories available on FPGA to
% avoid Block RAM port contention.

% \subsection{Memory allocation}
% Accelerators, whether they are GPU, FPGA or ASIC, all benefit from memory allocation
% optimizations. We present some of the existing domain-specific allocation techniques.

% \subsubsection{Access coalescing by dynamic program analysis}
% The previous optimizations rely on static program analysis. 
% It is also possible to obtain coalesced accesses by altering the program behavior
% using information collected at runtime. Works such as \cite{Fauzia_2015} 
% applying to GPU platforms alter the thread distribution on GPU when it leads 
% to coalesced accesses. This however has a limited applicability on FPGAs
% as the control logic is not updatable at runtime by the accelerator itself.

\section{Conclusion}
\label{sec:conclusion}

%Our work introduces a global memory layout for FPGA accelerators.
%It minimizes both the amount of memory transactions by optimal coalescing, and the 
%olume of transferred data by using careful static analysis.

This work gives a twofold contribution: a compression-friendly, contiguous
data layout, and an automated adaptation of FPGA accelerators to use this
layout thanks to polyhedral compilation tools. Thanks to the compression 
and contiguity, we can automatically reduce the number of I/O cycles spent
by the accelerator.

%In this work, we introduce a novel global memory layout for FPGA accelerators that maximises contiguity of the accessed data under constraint of irredundancy. We evaluate our approach against access patterns derived from the original layout and show that a MARS-based partitonning of the data allow a substantial increase of throughput and decrease of transfer time at the cost of a minimal resource increase.

\bibliographystyle{ACM-Reference-Format}
\bibliography{refs}

%%% -*-BibTeX-*-
%%% Do NOT edit. File created by BibTeX with style
%%% ACM-Reference-Format-Journals [18-Jan-2012].

\begin{thebibliography}{30}

%%% ====================================================================
%%% NOTE TO THE USER: you can override these defaults by providing
%%% customized versions of any of these macros before the \bibliography
%%% command.  Each of them MUST provide its own final punctuation,
%%% except for \shownote{}, \showDOI{}, and \showURL{}.  The latter two
%%% do not use final punctuation, in order to avoid confusing it with
%%% the Web address.
%%%
%%% To suppress output of a particular field, define its macro to expand
%%% to an empty string, or better, \unskip, like this:
%%%
%%% \newcommand{\showDOI}[1]{\unskip}   % LaTeX syntax
%%%
%%% \def \showDOI #1{\unskip}           % plain TeX syntax
%%%
%%% ====================================================================

\ifx \showCODEN    \undefined \def \showCODEN     #1{\unskip}     \fi
\ifx \showDOI      \undefined \def \showDOI       #1{#1}\fi
\ifx \showISBNx    \undefined \def \showISBNx     #1{\unskip}     \fi
\ifx \showISBNxiii \undefined \def \showISBNxiii  #1{\unskip}     \fi
\ifx \showISSN     \undefined \def \showISSN      #1{\unskip}     \fi
\ifx \showLCCN     \undefined \def \showLCCN      #1{\unskip}     \fi
\ifx \shownote     \undefined \def \shownote      #1{#1}          \fi
\ifx \showarticletitle \undefined \def \showarticletitle #1{#1}   \fi
\ifx \showURL      \undefined \def \showURL       {\relax}        \fi
% The following commands are used for tagged output and should be
% invisible to TeX
\providecommand\bibfield[2]{#2}
\providecommand\bibinfo[2]{#2}
\providecommand\natexlab[1]{#1}
\providecommand\showeprint[2][]{arXiv:#2}

\bibitem[Aarrestad et~al\mbox{.}(2021)]%
        {Aarrestad_2021}
\bibfield{author}{\bibinfo{person}{Thea Aarrestad}, \bibinfo{person}{Vladimir
  Loncar}, \bibinfo{person}{Nicol{\`{o}} Ghielmetti}, \bibinfo{person}{Maurizio
  Pierini}, \bibinfo{person}{Sioni Summers}, \bibinfo{person}{Jennifer
  Ngadiuba}, \bibinfo{person}{Christoffer Petersson}, \bibinfo{person}{Hampus
  Linander}, \bibinfo{person}{Yutaro Iiyama}, \bibinfo{person}{Giuseppe~Di
  Guglielmo}, \bibinfo{person}{Javier Duarte}, \bibinfo{person}{Philip Harris},
  \bibinfo{person}{Dylan Rankin}, \bibinfo{person}{Sergo Jindariani},
  \bibinfo{person}{Kevin Pedro}, \bibinfo{person}{Nhan Tran},
  \bibinfo{person}{Mia Liu}, \bibinfo{person}{Edward Kreinar},
  \bibinfo{person}{Zhenbin Wu}, {and} \bibinfo{person}{Duc Hoang}.}
  \bibinfo{year}{2021}\natexlab{}.
\newblock \showarticletitle{Fast convolutional neural networks on {FPGAs} with
  hls4ml}.
\newblock \bibinfo{journal}{\emph{Machine Learning: Science and Technology}}
  \bibinfo{volume}{2}, \bibinfo{number}{4} (\bibinfo{date}{jul}
  \bibinfo{year}{2021}), \bibinfo{pages}{045015}.
\newblock
\urldef\tempurl%
\url{https://doi.org/10.1088/2632-2153/ac0ea1}
\showDOI{\tempurl}


\bibitem[Bondhugula(2013)]%
        {Bondhugula_2013}
\bibfield{author}{\bibinfo{person}{Uday Bondhugula}.}
  \bibinfo{year}{2013}\natexlab{}.
\newblock \showarticletitle{Compiling Affine Loop Nests for Distributed-Memory
  Parallel Architectures}. In \bibinfo{booktitle}{\emph{Proceedings of the
  International Conference on High Performance Computing, Networking, Storage
  and Analysis}}. \bibinfo{publisher}{{ACM}}.
\newblock
\urldef\tempurl%
\url{https://doi.org/10.1145/2503210.2503289}
\showDOI{\tempurl}


\bibitem[Chi et~al\mbox{.}(2018)]%
        {Chi_2018}
\bibfield{author}{\bibinfo{person}{Yuze Chi}, \bibinfo{person}{Jason Cong},
  \bibinfo{person}{Peng Wei}, {and} \bibinfo{person}{Peipei Zhou}.}
  \bibinfo{year}{2018}\natexlab{}.
\newblock \showarticletitle{SODA: Stencil with optimized dataflow
  architecture}. In \bibinfo{booktitle}{\emph{2018 IEEE/ACM International
  Conference on Computer-Aided Design (ICCAD)}}. IEEE, \bibinfo{pages}{1--8}.
\newblock
\urldef\tempurl%
\url{https://doi.org/10.1145/3240765.3240850}
\showDOI{\tempurl}


\bibitem[Choi and Cong(2018)]%
        {choi2018hls}
\bibfield{author}{\bibinfo{person}{Young-kyu Choi} {and} \bibinfo{person}{Jason
  Cong}.} \bibinfo{year}{2018}\natexlab{}.
\newblock \showarticletitle{HLS-based optimization and design space exploration
  for applications with variable loop bounds}. In
  \bibinfo{booktitle}{\emph{2018 IEEE/ACM International Conference on
  Computer-Aided Design (ICCAD)}}. IEEE, \bibinfo{pages}{1--8}.
\newblock


\bibitem[Cong et~al\mbox{.}(2016)]%
        {cong2016source}
\bibfield{author}{\bibinfo{person}{Jason Cong}, \bibinfo{person}{Muhuan Huang},
  \bibinfo{person}{Peichen Pan}, \bibinfo{person}{Yuxin Wang}, {and}
  \bibinfo{person}{Peng Zhang}.} \bibinfo{year}{2016}\natexlab{}.
\newblock \showarticletitle{Source-to-source optimization for HLS}.
\newblock \bibinfo{journal}{\emph{FPGAs for Software Programmers}}
  (\bibinfo{year}{2016}), \bibinfo{pages}{137--163}.
\newblock


\bibitem[Cummiskey et~al\mbox{.}(1973)]%
        {Cummiskey_1973}
\bibfield{author}{\bibinfo{person}{P Cummiskey}, \bibinfo{person}{Nikil
  S.~Jayant}, {and} \bibinfo{person}{James L.~Flanagan}.}
  \bibinfo{year}{1973}\natexlab{}.
\newblock \showarticletitle{Adaptive Quantization in Differential PCM Coding of
  Speech}.
\newblock \bibinfo{journal}{\emph{Bell System Technical Journal}}
  \bibinfo{volume}{52} (\bibinfo{date}{09} \bibinfo{year}{1973}).
\newblock
\urldef\tempurl%
\url{https://doi.org/10.1002/j.1538-7305.1973.tb02007.x}
\showDOI{\tempurl}


\bibitem[Dathathri et~al\mbox{.}(2013)]%
        {Dathathri_2013}
\bibfield{author}{\bibinfo{person}{Roshan Dathathri}, \bibinfo{person}{Chandan
  Reddy}, \bibinfo{person}{Thejas Ramashekar}, {and} \bibinfo{person}{Uday
  Bondhugula}.} \bibinfo{year}{2013}\natexlab{}.
\newblock \showarticletitle{Generating Efficient Data Movement Code for
  Heterogeneous Architectures with Distributed-Memory}. In
  \bibinfo{booktitle}{\emph{Proceedings of the 22nd International Conference on
  Parallel Architectures and Compilation Techniques}}.
  \bibinfo{publisher}{{IEEE}}.
\newblock
\urldef\tempurl%
\url{https://doi.org/10.1109/PACT.2013.6618833}
\showDOI{\tempurl}


\bibitem[Du et~al\mbox{.}(2022)]%
        {Du_2022}
\bibfield{author}{\bibinfo{person}{Yixiao Du}, \bibinfo{person}{Yuwei Hu},
  \bibinfo{person}{Zhongchun Zhou}, {and} \bibinfo{person}{Zhiru Zhang}.}
  \bibinfo{year}{2022}\natexlab{}.
\newblock \showarticletitle{High-Performance Sparse Linear Algebra on
  {HBM}-Equipped {FPGAs} Using {HLS}}. In \bibinfo{booktitle}{\emph{Proceedings
  of the 2022 {ACM}/{SIGDA} International Symposium on Field-Programmable Gate
  Arrays}}. \bibinfo{publisher}{{ACM}}.
\newblock
\urldef\tempurl%
\url{https://doi.org/10.1145/3490422.3502368}
\showDOI{\tempurl}


\bibitem[Ferry et~al\mbox{.}(2023)]%
        {Ferry_2023}
\bibfield{author}{\bibinfo{person}{Corentin Ferry}, \bibinfo{person}{Steven
  Derrien}, {and} \bibinfo{person}{Sanjay Rajopadhye}.}
  \bibinfo{year}{2023}\natexlab{}.
\newblock \showarticletitle{Maximal Atomic irRedundant Sets: a Usage-based
  Dataflow Partitioning Algorithm}. In \bibinfo{booktitle}{\emph{13th
  International Workshop on Polyhedral Compilation Techniques (IMPACT’23)}}.
\newblock


\bibitem[Ferry et~al\mbox{.}(2022)]%
        {Ferry_2022}
\bibfield{author}{\bibinfo{person}{Corentin Ferry}, \bibinfo{person}{Tomofumi
  Yuki}, \bibinfo{person}{Steven Derrien}, {and} \bibinfo{person}{Sanjay
  Rajopadhye}.} \bibinfo{year}{2022}\natexlab{}.
\newblock \showarticletitle{Increasing {FPGA} Accelerators Memory Bandwidth
  with a Burst-Friendly Memory Layout}.
\newblock \bibinfo{journal}{\emph{{IEEE} Transactions on Computer-Aided Design
  of Integrated Circuits and Systems}} (\bibinfo{year}{2022}),
  \bibinfo{pages}{1--1}.
\newblock
\urldef\tempurl%
\url{https://doi.org/10.1109/tcad.2022.3201494}
\showDOI{\tempurl}


\bibitem[Guan et~al\mbox{.}(2017)]%
        {Guan_2017}
\bibfield{author}{\bibinfo{person}{Yijin Guan}, \bibinfo{person}{Ningyi Xu},
  \bibinfo{person}{Chen Zhang}, \bibinfo{person}{Zhihang Yuan}, {and}
  \bibinfo{person}{Jason Cong}.} \bibinfo{year}{2017}\natexlab{}.
\newblock \showarticletitle{Using Data Compression for Optimizing {FPGA}-Based
  Convolutional Neural Network Accelerators}.
\newblock In \bibinfo{booktitle}{\emph{Lecture Notes in Computer Science}}.
  \bibinfo{publisher}{Springer International Publishing},
  \bibinfo{pages}{14--26}.
\newblock
\urldef\tempurl%
\url{https://doi.org/10.1007/978-3-319-67952-5_2}
\showDOI{\tempurl}


\bibitem[Horro et~al\mbox{.}(2023)]%
        {Horro_2022}
\bibfield{author}{\bibinfo{person}{Marcos Horro},
  \bibinfo{person}{Louis-No\"{e}l Pouchet}, \bibinfo{person}{Gabriel
  Rodr\'{\i}guez}, {and} \bibinfo{person}{Juan Touri\~{n}o}.}
  \bibinfo{year}{2023}\natexlab{}.
\newblock \showarticletitle{Custom High-Performance Vector Code Generation for
  Data-Specific Sparse Computations}. In \bibinfo{booktitle}{\emph{Proceedings
  of the International Conference on Parallel Architectures and Compilation
  Techniques}} (Chicago, Illinois) \emph{(\bibinfo{series}{PACT '22})}.
  \bibinfo{publisher}{Association for Computing Machinery},
  \bibinfo{address}{New York, NY, USA}, \bibinfo{pages}{160–171}.
\newblock
\showISBNx{9781450398688}
\urldef\tempurl%
\url{https://doi.org/10.1145/3559009.3569668}
\showDOI{\tempurl}


\bibitem[Li et~al\mbox{.}(2023)]%
        {Li_2023}
\bibfield{author}{\bibinfo{person}{Shiqing Li}, \bibinfo{person}{Di Liu}, {and}
  \bibinfo{person}{Weichen Liu}.} \bibinfo{year}{2023}\natexlab{}.
\newblock \showarticletitle{Efficient {FPGA}-based Sparse Matrix-Vector
  Multiplication with Data Reuse-aware Compression}.
\newblock \bibinfo{journal}{\emph{{IEEE} Transactions on Computer-Aided Design
  of Integrated Circuits and Systems}} (\bibinfo{year}{2023}),
  \bibinfo{pages}{1--1}.
\newblock
\urldef\tempurl%
\url{https://doi.org/10.1109/tcad.2023.3281715}
\showDOI{\tempurl}


\bibitem[Liu et~al\mbox{.}(2015)]%
        {liu2015offline}
\bibfield{author}{\bibinfo{person}{Junyi Liu}, \bibinfo{person}{Samuel
  Bayliss}, {and} \bibinfo{person}{George~A Constantinides}.}
  \bibinfo{year}{2015}\natexlab{}.
\newblock \showarticletitle{Offline synthesis of online dependence testing:
  Parametric loop pipelining for HLS}. In \bibinfo{booktitle}{\emph{2015 IEEE
  23rd Annual International Symposium on Field-Programmable Custom Computing
  Machines}}. IEEE, \bibinfo{pages}{159--162}.
\newblock


\bibitem[Liu et~al\mbox{.}(2017)]%
        {liu2017polyhedral}
\bibfield{author}{\bibinfo{person}{Junyi Liu}, \bibinfo{person}{John
  Wickerson}, \bibinfo{person}{Samuel Bayliss}, {and} \bibinfo{person}{George~A
  Constantinides}.} \bibinfo{year}{2017}\natexlab{}.
\newblock \showarticletitle{Polyhedral-based dynamic loop pipelining for
  high-level synthesis}.
\newblock \bibinfo{journal}{\emph{IEEE Transactions on Computer-Aided Design of
  Integrated Circuits and Systems}} \bibinfo{volume}{37}, \bibinfo{number}{9}
  (\bibinfo{year}{2017}), \bibinfo{pages}{1802--1815}.
\newblock


\bibitem[Lo et~al\mbox{.}(2023)]%
        {lo2023hmlib}
\bibfield{author}{\bibinfo{person}{Michael Lo}, \bibinfo{person}{Young-kyu
  Choi}, \bibinfo{person}{Weikang Qiao}, \bibinfo{person}{Mau-Chung~Frank
  Chang}, {and} \bibinfo{person}{Jason Cong}.} \bibinfo{year}{2023}\natexlab{}.
\newblock \showarticletitle{HMLib: Efficient Data Transfer for HLS Using Host
  Memory}. In \bibinfo{booktitle}{\emph{Proceedings of the 2023 ACM/SIGDA
  International Symposium on Field Programmable Gate Arrays}}.
  \bibinfo{pages}{50--50}.
\newblock


\bibitem[Mayer et~al\mbox{.}(2023)]%
        {mayer2023employing}
\bibfield{author}{\bibinfo{person}{Florian Mayer}, \bibinfo{person}{Julian
  Brandner}, {and} \bibinfo{person}{Michael Philippsen}.}
  \bibinfo{year}{2023}\natexlab{}.
\newblock \showarticletitle{Employing Polyhedral Methods to Reduce Data
  Movement in FPGA Stencil Codes}. In \bibinfo{booktitle}{\emph{International
  Workshop on Languages and Compilers for Parallel Computing}}. Springer,
  \bibinfo{pages}{47--63}.
\newblock


\bibitem[Nakahara et~al\mbox{.}(2020)]%
        {Nakahara_2020}
\bibfield{author}{\bibinfo{person}{Hiroki Nakahara}, \bibinfo{person}{Zhiqiang
  Que}, {and} \bibinfo{person}{Wayne Luk}.} \bibinfo{year}{2020}\natexlab{}.
\newblock \showarticletitle{High-Throughput Convolutional Neural Network on an
  FPGA by Customized JPEG Compression}. In \bibinfo{booktitle}{\emph{2020 IEEE
  28th Annual International Symposium on Field-Programmable Custom Computing
  Machines (FCCM)}}. \bibinfo{pages}{1--9}.
\newblock
\urldef\tempurl%
\url{https://doi.org/10.1109/FCCM48280.2020.00010}
\showDOI{\tempurl}


\bibitem[Ozturk et~al\mbox{.}(2009)]%
        {Ozturk_2009}
\bibfield{author}{\bibinfo{person}{O. Ozturk}, \bibinfo{person}{M. Kandemir},
  {and} \bibinfo{person}{M.J. Irwin}.} \bibinfo{year}{2009}\natexlab{}.
\newblock \showarticletitle{Using Data Compression for Increasing Memory System
  Utilization}.
\newblock \bibinfo{journal}{\emph{{IEEE} Transactions on Computer-Aided Design
  of Integrated Circuits and Systems}} \bibinfo{volume}{28},
  \bibinfo{number}{6} (\bibinfo{date}{jun} \bibinfo{year}{2009}),
  \bibinfo{pages}{901--914}.
\newblock
\urldef\tempurl%
\url{https://doi.org/10.1109/tcad.2009.2017430}
\showDOI{\tempurl}


\bibitem[Pouchet and Yuki(2016)]%
        {Polybench}
\bibfield{author}{\bibinfo{person}{Louis-No{\"e}l Pouchet} {and}
  \bibinfo{person}{Tomofumi Yuki}.} \bibinfo{year}{2016}\natexlab{}.
\newblock \bibinfo{title}{PolyBench/C 4.2.1}.
\newblock
\newblock
\urldef\tempurl%
\url{http://polybench.sf.net}
\showURL{%
\tempurl}


\bibitem[Pouchet et~al\mbox{.}(2013)]%
        {Pouchet_2013}
\bibfield{author}{\bibinfo{person}{Louis-Noel Pouchet}, \bibinfo{person}{Peng
  Zhang}, \bibinfo{person}{P. Sadayappan}, {and} \bibinfo{person}{Jason Cong}.}
  \bibinfo{year}{2013}\natexlab{}.
\newblock \showarticletitle{Polyhedral-based data reuse optimization for
  configurable computing}. In \bibinfo{booktitle}{\emph{Proceedings of the
  {ACM}/{SIGDA} international symposium on Field programmable gate arrays -
  {FPGA} {\textquotesingle}13}}. \bibinfo{publisher}{{ACM} Press}.
\newblock
\urldef\tempurl%
\url{https://doi.org/10.1145/2435264.2435273}
\showDOI{\tempurl}


\bibitem[Santos and Cardoso(2020)]%
        {santos2020automatic}
\bibfield{author}{\bibinfo{person}{Tiago Santos} {and}
  \bibinfo{person}{Jo{\~a}o~MP Cardoso}.} \bibinfo{year}{2020}\natexlab{}.
\newblock \showarticletitle{Automatic selection and insertion of hls directives
  via a source-to-source compiler}. In \bibinfo{booktitle}{\emph{2020
  International Conference on Field-Programmable Technology (ICFPT)}}. IEEE,
  \bibinfo{pages}{227--232}.
\newblock


\bibitem[Sardashti et~al\mbox{.}(2016)]%
        {Sardashti_2016}
\bibfield{author}{\bibinfo{person}{Somayeh Sardashti}, \bibinfo{person}{Andre
  Seznec}, {and} \bibinfo{person}{David~A. Wood}.}
  \bibinfo{year}{2016}\natexlab{}.
\newblock \showarticletitle{Yet Another Compressed Cache: A Low-Cost Yet
  Effective Compressed Cache}.
\newblock \bibinfo{journal}{\emph{ACM Trans. Archit. Code Optim.}}
  \bibinfo{volume}{13}, \bibinfo{number}{3}, Article \bibinfo{articleno}{27}
  (\bibinfo{date}{Sept.} \bibinfo{year}{2016}), \bibinfo{numpages}{25}~pages.
\newblock
\showISSN{1544-3566}
\urldef\tempurl%
\url{https://doi.org/10.1145/2976740}
\showDOI{\tempurl}


\bibitem[Skodras et~al\mbox{.}(2001)]%
        {Skodras_2001}
\bibfield{author}{\bibinfo{person}{A. Skodras}, \bibinfo{person}{C.
  Christopoulos}, {and} \bibinfo{person}{T. Ebrahimi}.}
  \bibinfo{year}{2001}\natexlab{}.
\newblock \showarticletitle{The JPEG 2000 still image compression standard}.
\newblock \bibinfo{journal}{\emph{IEEE Signal Processing Magazine}}
  \bibinfo{volume}{18}, \bibinfo{number}{5} (\bibinfo{year}{2001}),
  \bibinfo{pages}{36--58}.
\newblock
\urldef\tempurl%
\url{https://doi.org/10.1109/79.952804}
\showDOI{\tempurl}


\bibitem[Sohrabizadeh et~al\mbox{.}(2022)]%
        {sohrabizadeh2022autodse}
\bibfield{author}{\bibinfo{person}{Atefeh Sohrabizadeh},
  \bibinfo{person}{Cody~Hao Yu}, \bibinfo{person}{Min Gao}, {and}
  \bibinfo{person}{Jason Cong}.} \bibinfo{year}{2022}\natexlab{}.
\newblock \showarticletitle{AutoDSE: Enabling software programmers to design
  efficient FPGA accelerators}.
\newblock \bibinfo{journal}{\emph{ACM Transactions on Design Automation of
  Electronic Systems (TODAES)}} \bibinfo{volume}{27}, \bibinfo{number}{4}
  (\bibinfo{year}{2022}), \bibinfo{pages}{1--27}.
\newblock


\bibitem[Tian et~al\mbox{.}(2020)]%
        {Tian_2020}
\bibfield{author}{\bibinfo{person}{Teng Tian}, \bibinfo{person}{Xi Jin},
  \bibinfo{person}{Letian Zhao}, \bibinfo{person}{Xiaotian Wang},
  \bibinfo{person}{Jie Wang}, {and} \bibinfo{person}{Wei Wu}.}
  \bibinfo{year}{2020}\natexlab{}.
\newblock \showarticletitle{Exploration of Memory Access Optimization for
  FPGA-based 3D CNN Accelerator}. In \bibinfo{booktitle}{\emph{2020 Design,
  Automation \& Test in Europe Conference \& Exhibition (DATE)}}.
  \bibinfo{pages}{1650--1655}.
\newblock
\urldef\tempurl%
\url{https://doi.org/10.23919/DATE48585.2020.9116376}
\showDOI{\tempurl}


\bibitem[Verdoolaege(2010)]%
        {Verdoolaege_2016}
\bibfield{author}{\bibinfo{person}{Sven Verdoolaege}.}
  \bibinfo{year}{2010}\natexlab{}.
\newblock \showarticletitle{isl: An Integer Set Library for the Polyhedral
  Model}. In \bibinfo{booktitle}{\emph{Mathematical Software -- ICMS 2010}},
  \bibfield{editor}{\bibinfo{person}{Komei Fukuda}, \bibinfo{person}{Joris
  van~der Hoeven}, \bibinfo{person}{Michael Joswig}, {and}
  \bibinfo{person}{Nobuki Takayama}} (Eds.). \bibinfo{publisher}{Springer
  Berlin Heidelberg}, \bibinfo{address}{Berlin, Heidelberg},
  \bibinfo{pages}{299--302}.
\newblock
\showISBNx{978-3-642-15582-6}


\bibitem[Ye et~al\mbox{.}(2022)]%
        {ye2022scalehls}
\bibfield{author}{\bibinfo{person}{Hanchen Ye}, \bibinfo{person}{Cong Hao},
  \bibinfo{person}{Jianyi Cheng}, \bibinfo{person}{Hyunmin Jeong},
  \bibinfo{person}{Jack Huang}, \bibinfo{person}{Stephen Neuendorffer}, {and}
  \bibinfo{person}{Deming Chen}.} \bibinfo{year}{2022}\natexlab{}.
\newblock \showarticletitle{Scalehls: A new scalable high-level synthesis
  framework on multi-level intermediate representation}. In
  \bibinfo{booktitle}{\emph{2022 IEEE International Symposium on
  High-Performance Computer Architecture (HPCA)}}. IEEE,
  \bibinfo{pages}{741--755}.
\newblock


\bibitem[Zhao et~al\mbox{.}(2022)]%
        {zhao2022polsca}
\bibfield{author}{\bibinfo{person}{Ruizhe Zhao}, \bibinfo{person}{Jianyi
  Cheng}, \bibinfo{person}{Wayne Luk}, {and} \bibinfo{person}{George~A
  Constantinides}.} \bibinfo{year}{2022}\natexlab{}.
\newblock \showarticletitle{POLSCA: Polyhedral High-Level Synthesis with
  Compiler Transformations}. In \bibinfo{booktitle}{\emph{2022 32nd
  International Conference on Field-Programmable Logic and Applications
  (FPL)}}. IEEE, \bibinfo{pages}{235--242}.
\newblock


\bibitem[Zhao et~al\mbox{.}(2021)]%
        {Zhao_2021}
\bibfield{author}{\bibinfo{person}{Tuowen Zhao}, \bibinfo{person}{Mary Hall},
  \bibinfo{person}{Hans Johansen}, {and} \bibinfo{person}{Samuel Williams}.}
  \bibinfo{year}{2021}\natexlab{}.
\newblock \showarticletitle{Improving communication by optimizing on-node data
  movement with data layout}. In \bibinfo{booktitle}{\emph{Proceedings of the
  26th {ACM} {SIGPLAN} Symposium on Principles and Practice of Parallel
  Programming}}. \bibinfo{publisher}{{ACM}}.
\newblock
\urldef\tempurl%
\url{https://doi.org/10.1145/3437801.3441598}
\showDOI{\tempurl}


\end{thebibliography}

\end{document}